\documentclass[%
 reprint,
 amsmath,amssymb,
 aps,
]{revtex4-2}

\usepackage{verbatim}
\usepackage{graphicx}
\usepackage{dcolumn}
\usepackage{bm}
\usepackage[export]{adjustbox}
\usepackage{placeins}
\usepackage{hyperref}

\begin{document}


\title{A Beehive Haloscope for High-mass Axion Dark Matter}

\author{Matthew O. Withers}
\affiliation{%
 Stanford University, Stanford, CA 94305, USA
}%
\author{Chao-Lin Kuo}
\affiliation{%
 Stanford University, Stanford, CA 94305, USA
}%
\affiliation{%
 SLAC National Accelerator Laboratory, 2575 Sand Hill Road,
Menlo Park, CA 94025, USA
}

\date{\today}

\begin{abstract}
We propose a new haloscope geometry that can arbitrarily increase the resonator volume for a given target axion mass.  This geometry consists of closely packed, {\em overlapping} coaxial cavities operating as a single resonator.  While the resonant frequency is still determined by the dimensions of the individual ``cells,'' the strong interactions between the cells encourage the entire ``beehive'' to oscillate {\em in phase}, a phenomenon expected of tightly coupled harmonic oscillators.  This synchronization behavior allows the construction of a singly connected large-volume resonator at the high frequency by simply increasing the number of the cells. Using direct numerical simulations, we verify the existence of a global eigenmode that has a high ($40\%$) form factor in a 169-element beehive resonator.  The resonant frequency of the eigenmode is tunable by moving the center rods laterally in unison.  The form factor is very tolerant to dimensional deviations and misalignment, as a result of mode hybridization due to strong coupling. The beehive haloscope inherits many appealing properties from the conventional coaxial cavity: a high quality factor, compatibility with a solenoid magnet, ease of fabrication, tuning, and coupling. We argue that this geometry is an excellent candidate for high-mass axion searches covering the post-inflationary parameter space ($>$5 GHz).    
\end{abstract}

\maketitle

\section{Introduction}

\subsection{Motivation}
 
Current astronomical and cosmological observations, including measurements of galaxy rotation curves \cite{1970ApJ...159..379R}, gravitational lensing \cite{DES3}, cosmic microwave background (CMB) anisotropies \cite{Planck2020}, and galaxy clusters \cite{bullet}, indicate that dark matter constitutes $\sim$85\% of the mass of the universe. Loosely characterized as feebly interacting, nonrelativistic, and nonbaryonic, the particle identity of dark matter is unknown, despite the pivotal role it must have played in the evolution of the universe under the Lambda cold dark matter ($\Lambda$CDM) model of cosmology. 

Several promising explanations of the nature and origins of dark matter are currently under investigation. One was initially proposed by Peccei and Quinn (PQ) to solve the strong-CP problem in quantum chromodynamics (QCD). The PQ mechanism explains the absence of CP violation in neutrons, despite the presence of a CP-violating term in the QCD Lagrangian. Their solution introduced an additional global U(1) symmetry to QCD, which, when broken, creates a pseudo Nambu-Goldstone boson, dubbed the axion \cite{PecciQuinn1977, Wilczek1978, Weinberg1978,abbott83,preskill83,dine83}.

Despite their weakly-interacting nature, axions exhibit a coupling to photons via the constant $g_{a \gamma \gamma}$, making them, in principle, experimentally detectable. The two prevailing models for this coupling, the Kim-Shifman-Vainshtein-Zakharov (KSVZ) \cite{kim,shifman} and Dine-Fischler-Srednicki-Zhitnitsky (DFSZ) \cite{dine,Zhitnitsky:1980tq} models, both predict an extremely small, frequency-dependent value for $g_{a \gamma \gamma}$ and thus set benchmarks for any experiment seeking to detect axion to photon conversion.

Under strong magnetic fields, axions convert to photons via the inverse Primakoff effect (see \cite{marsh} and references therein). The frequency of the converted photons is proportional to the unknown axion mass. One class of experiments, based on Sikivie's axion haloscope, seeks to detect this conversion by placing a tunable resonating cavity inside a strong magnetic field \cite{sikivie1}. Photons produced via the inverse Primakoff effect will resonate inside the cavity; this signal can be transferred to an rf chain via a coupled antenna. By tuning the cavity's resonant frequency, one can search for the existence of axions at different masses, thus probing the coupling-mass axion parameter space. To date, only cavity-based haloscopes have probed this space at KSVZ and DFSZ sensitivity \cite{admx18,admx19,ADMX:2021nhd,capp}.

Resonator-based haloscopes have a scan speed that scales as $\frac{d\nu}{dt} \propto B_0^4 C^2 V^2 Q$\cite{sikivie1}, where $\nu$ is the detected photon frequency, $B_0$ is the strength of the magnetic field, $V$ is the cavity volume, $Q$ is the cavity quality factor, and $C$ is the form factor. Current efforts seek to expand KSVZ and DSVZ-level sensitivity from around 1 GHz to cover the post-inflationary scenario, which approximately falls in the mass range of 20-200 $\mu$eV (5-50 GHz) \cite{marsh,Borsanyi_16,Klaer_2017,Buschmann2022}. As frequency increases, a scaled conventional cylindrical haloscope cavity will experience a volume reduction $V \propto \nu^{-3}$, which implies a scan rate reduction $\frac{d\nu}{dt} \propto \nu^{-6}$. This decrease, coupled with increased noise from cryogenic amplifiers within the GHz band and a decreasing quality factor, severely limits the scan speed of these searches. 

The halsocope scheme has been modified to increase the active volume at higher frequencies. One such approach is the introduction of multiple cavities (\textit{e.g.}, ADMX-EFR and its pathfinder ADMX-G2). This approach, while maintaining the simple cavity geometry, introduces challenges with maintaining phase synchronization between cavities and rf chain complexity. 

A second approach involves cavity geometries that fall into the thin-shell category. First proposed in \cite{Kuo_2020}, this type of geometry has a frequency-fixing dimension (which will be ``thin'' at high frequencies), while growing two other dimensions to achieve a high volume. This scheme reduces the active volume scaling to $V \propto \nu^{-1}$. A variety of overall shapes and tuning mechanisms have been studied \cite{Kuo_2021,rectangle,dyson}.

Alternatively, single-volume cavities with multiple cells or tuning elements have been explored \cite{jeong23,Simanovskaia_2021}. These designs feature demonstratively larger effective volume than a single cavity and fewer (often one single) readout ports than the number of cells.  For the proposed geometries to date, the tuning mechanism tends to involve many degrees of freedom. Moreover, there is not an obviously scalable implementation that can produce a detector volume $>$ 100 $\lambda^3$. 

In this paper, we seek to expand and integrate the multiple-cell and thin-shelled cavity schemes by proposing a geometry of closely packed, overlapping coaxial cavities. Closely resembling the arrangement of cells in a beehive, this approach has several advantages, including the ability to arbitrarily increase the active volume for a given target frequency range. Furthermore, the cells of the cavity operate as a single resonator due to strong coupling, guaranteeing in-phase operation without the need for software- or hardware- phase-locked loops. Another advantage of single resonator operation is shared coupling point(s), which significantly decreases the complexity of rf wiring, simplifying operations and reducing cost. Furthermore, the scheme maintains key haloscope features, including a high $Q$, compatibility with the field produced by a solenoid magnet, and ease of fabrication, and tuning.   

This paper is organized as follows. First, we explore the physics of strongly coupled oscillators and apply the results to overlapping cylindrical shells. Second, we propose our cavity geometry and describe its tuning mechanism. Third, we explore 2-dimensional finite element analyses (FEA) to verify the scalability, tunability, and resilience of the beehive scheme. Finally, we explore simulations of a prototype 3-cell cavity in 3-dimensions before comparing the expected performance of our approach to other experiments seeking to operate within a similar frequency range. 

\subsection{Theoretical Background: Multiple-Cell Cavity as Strongly Coupled Oscillators}
\label{sec:oscillators}

To begin our study, we model the interactions between the cells in a multiple-cell setup using coupled harmonic oscillators and COMSOL Multiphysics's \textit{RF Module} (COMSOL-RF).  We conclude from these studies that the coupling encourages the cells to oscillate collectively {\em in phase} despite individual variations in the natural frequency that exceed $1/Q$ fractionally. 
This newfound understanding leads to our proposal for the manifestly scalable beehive geometry, which is discussed in detail in the following sections. 
This collective behavior is superficially similar to the synchronization of clocks and metronomes first observed by Christiaan Huygens in the 17$^{\rm th}$ century and subsequently explained by many mathematicians (see \cite{metronome,strogatz93} and references therein).  Both types of synchronization are caused by coupling between multiple oscillators. However, the coupling strength between the cells in a cavity is much stronger than those between the Huygens' clocks. In what follows, we see that a strong coupling directly results in phase synchronization between the oscillators when one considers the components of the eigenvectors. 

\begin{figure}
    \centering
    \includegraphics[width=0.5\textwidth]{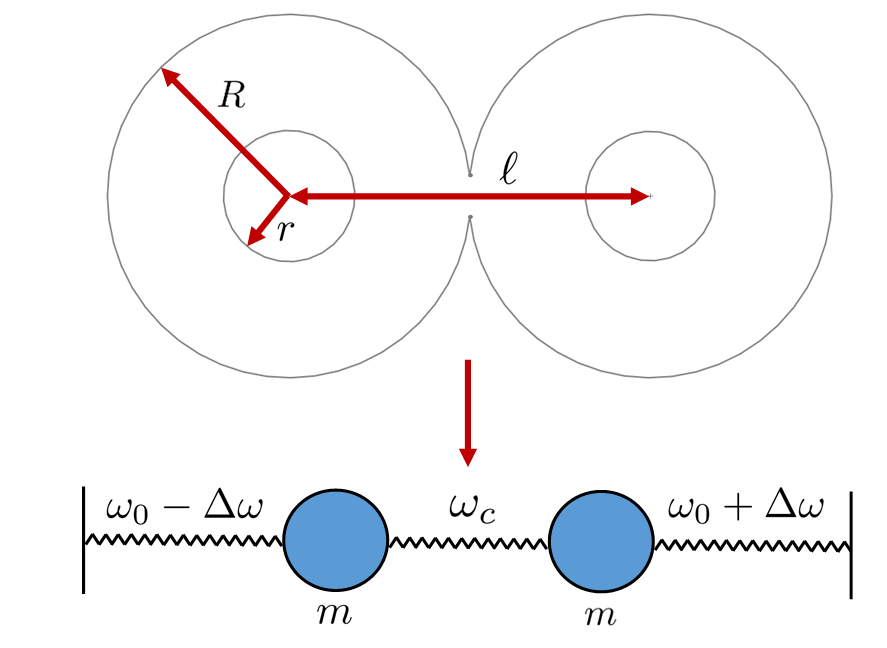}
    \caption{Spring model for two strongly-coupled coaxial cavities.}
    \label{fig:spring_model}
\end{figure}

The basic behavior of a two-cell cavity can be modeled by a pair of mass-spring oscillators (see Figure \ref{fig:spring_model}).
The natural frequencies of the oscillators are slightly detuned to $\omega_0-\Delta\omega$ and $\omega_0+\Delta\omega$, respectively. They are joined together with a third spring (with a spring constant $\kappa$). We denote the amount of frequency coupling by $\omega_c\equiv \sqrt{\kappa/m}$.  

Solving the characteristic equations leads to two distinct eigenfrequencies,  
\begin{equation}
\omega_{\pm}=\sqrt{\omega_0^2+\omega_c^2+\Delta\omega^2\mp\sqrt{4\omega_0^2\Delta\omega^2+\omega_c^4}}.
\label{eq:mode_hybridizations}
\end{equation}
These correspond to in-phase, or symmetric ($+$), and out-of-phase, or anti-symmetric ($-$) hybridizations. Note that $\omega_+$ is the lower of the two eigenfrequencies. Without detuning ($\Delta\omega=0$), the amount of splitting is given by $\Omega\equiv\omega_--\omega_+\sim \omega_c^2/\omega_0$, with $\omega_0\gg\omega_c$. Varying the detuning continuously leads to the well known avoided crossing phenomena (\textit{e.g.}, \cite{novotny10}).

For an axion haloscope, the form factor $C$ that quantifies the overlap between the electromagnetic mode and the external magnetic field (oriented in the $z$-direction) is given by
\begin{equation}
    C = \frac{(\int E_z dV)^2}{V\int E^2dV}.\label{eq:C_Definition}
\end{equation}
Note that when the volume integrals are performed over multiple cells, they become the sum of the volume integrals of the individual cells multiplied by the appropriate phase sensitive coefficients.  
The in-phase eigenvector $\boldsymbol{e_+}$ corresponding to $\omega_+$ contains the relevant coefficients for the axion-sensitive mode.  The eigenvector works out to be proportional to $[1,\sqrt{1+\zeta^2}-\zeta]$, where $\zeta\equiv 2\omega_0\Delta\omega/\omega_c^2\sim 2\Delta\omega/\Omega$.  The ``relative phase'' form factor $C_p$ can then be derived from Eq. \ref{eq:C_Definition} using the components of $\boldsymbol{e_+}$, 
$$
C_p=\frac{(1+\sqrt{1+\zeta^2}-\zeta)^2}{2[1+(\sqrt{1+\zeta^2}-\zeta)^2]}.
$$
In addition to the form factor for individual cells, this multiplicative factor $C_p$ tracks the fractional reduction in the effective volume due to dephasing.  Note that $C_p$ reduces to $\sim 1- \zeta^2/4$ for $\zeta \ll 1$ \footnote{$C_p\sim 0.5$ for $\zeta\gg 1$, meaning only half of the volume is active when the two cavities are completely detuned. A similar quantity vanishes in the case of anti-symmetric hybridization $\vec{e}_-$, corresponding to phase cancellation.}.

These results suggest that as long as the amount of detuning $\Delta \omega$ remains less than $\Omega$, we have $C_p\sim 1$ and there is only a small reduction in the form factor caused by dephasing. Thus, by introducing coupling between originally independent oscillators, the tolerance criterion for in-phase oscillation is relaxed from $\Delta \omega <\omega_0/Q$ (the line width) to $\Delta \omega < \Omega\sim \omega_c^2/\omega_0$ (the splitting). 

Increasing the amount of coupling $\omega_c$ makes the frequency splitting $\Omega$ larger and the system more tolerant to detuning. The synchronization phenomenon in this limit is much more elementary than Huygens synchronization, but has not been previously reported to our knowledge. 

\begin{figure}
    \includegraphics[width=0.48\textwidth,left]{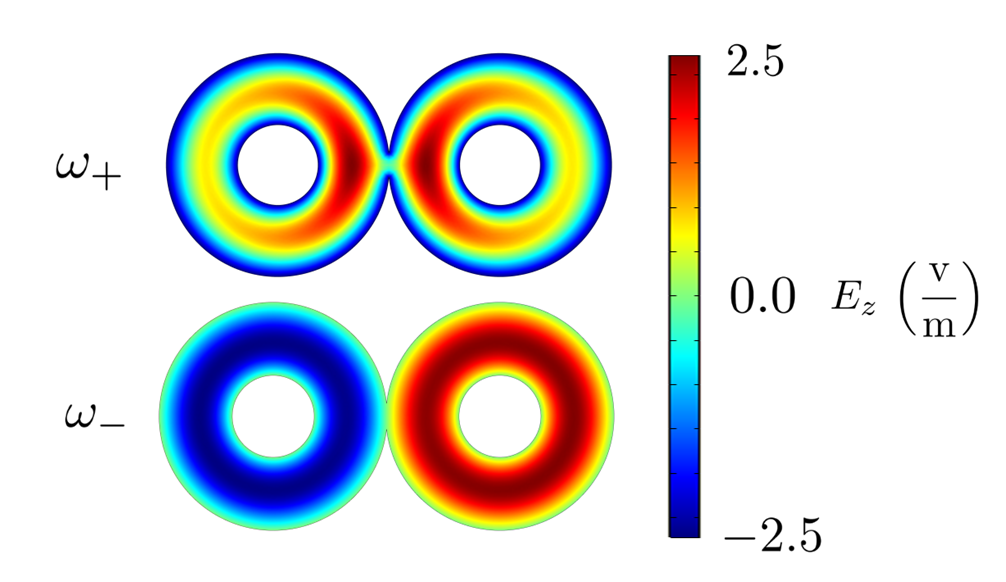}
    \caption{Symmetric $\omega_+$ and antisymmetric $\omega_-$ modes for a strongly-coupled two-cell cavity with $\ell=38.75$ mm, $R=19.5$ mm, and $r=7$ mm as simulated in COMSOL-RF. The simulated resonant frequency of $\omega_+/2\pi$ is 11.775 GHz, while the simulated resonant frequency of $\omega_-/2\pi$ is 11.846 GHz. }
    \label{fig:mode_examples}
\end{figure}

To confirm this behavior in a pair of overlapping cavities, we used COMSOL-RF to construct a simple two-dimensional cavity model consisting of two overlapping circular cells (upper sketch in Figure \ref{fig:spring_model}). We varied the coupling between the two cells by adjusting the amount of overlap (\textit{i.e.}, reducing the center-to-center distance $\ell$ between the cells for stronger coupling or increasing it for weaker coupling). In the center of the cells (radii $R$), we placed circular tuning rods and adjusted their radii $r$ to detune the system. For simplicity, we limited our analysis to situations where the rod radii varied antisymmetrically (\textit{i.e.}, $r_{left} = r \pm \delta r$ and $r_{right} = r \mp \delta r$). In such cases, our simulations revealed two modes of interest. The first is symmetric between the two cells (see upper color map in Figure \ref{fig:mode_examples}) and corresponds to $\omega_+$ from our theoretical analysis. The second is antisymmetric between the two cells (see lower color map in Figure \ref{fig:mode_examples}) and is equivalent to $\omega_-$.

\begin{figure}
    \includegraphics[left]{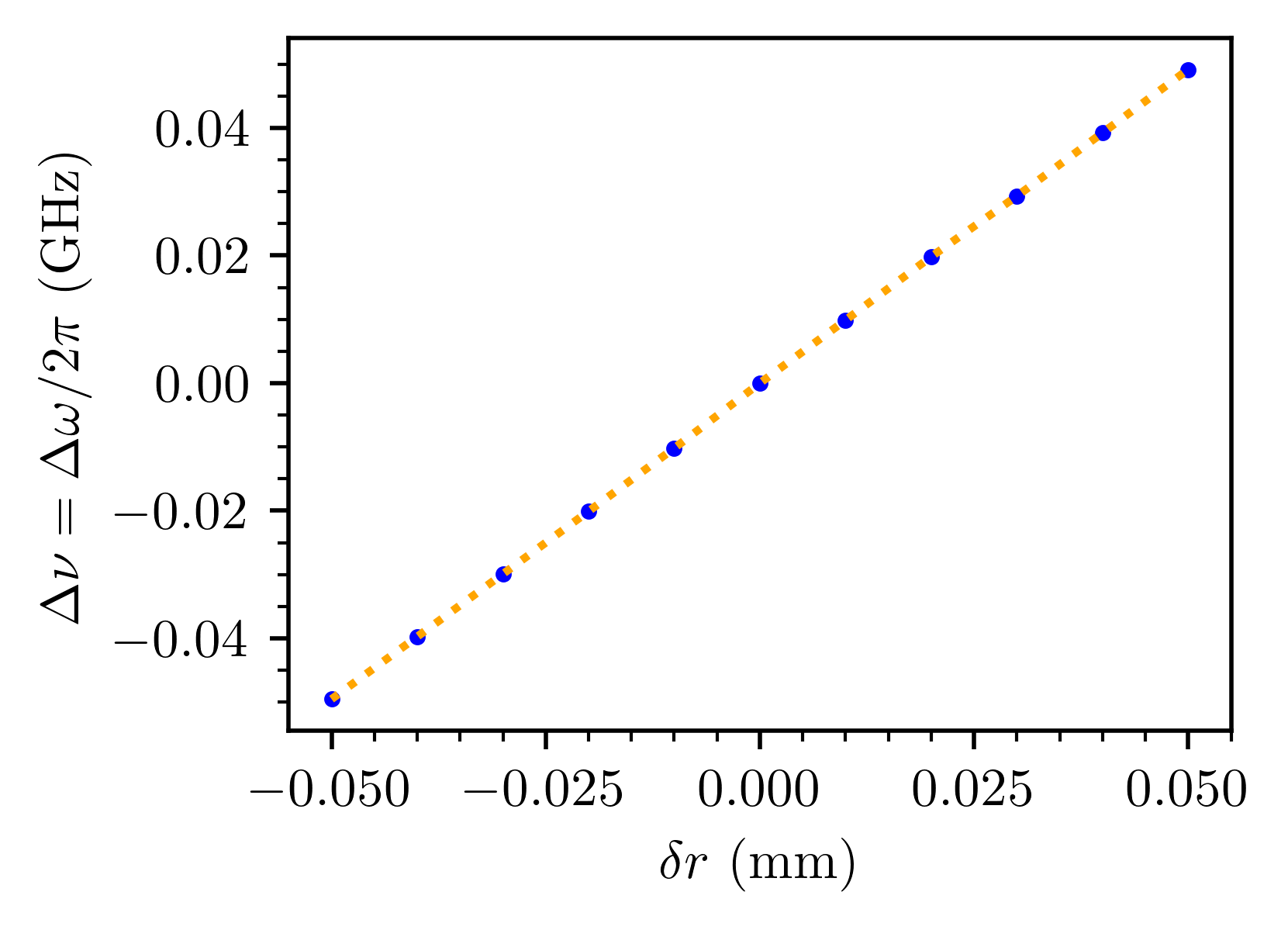}
    \caption{Tuning rod radius variation $\delta r$ versus simulated resonant frequency detuning $\Delta \nu = \Delta \omega / 2 \pi$ for the symmetric mode of a single-cell cavity (blue points). The cell radius was $R=19.5$ mm. We interpolate this data (orange dotted line) to determine the relationship betweeen $\delta r$ and $\Delta \nu$.}
    \label{fig:radius_freq_relationship}
\end{figure}

\begin{figure}
    \includegraphics[left]{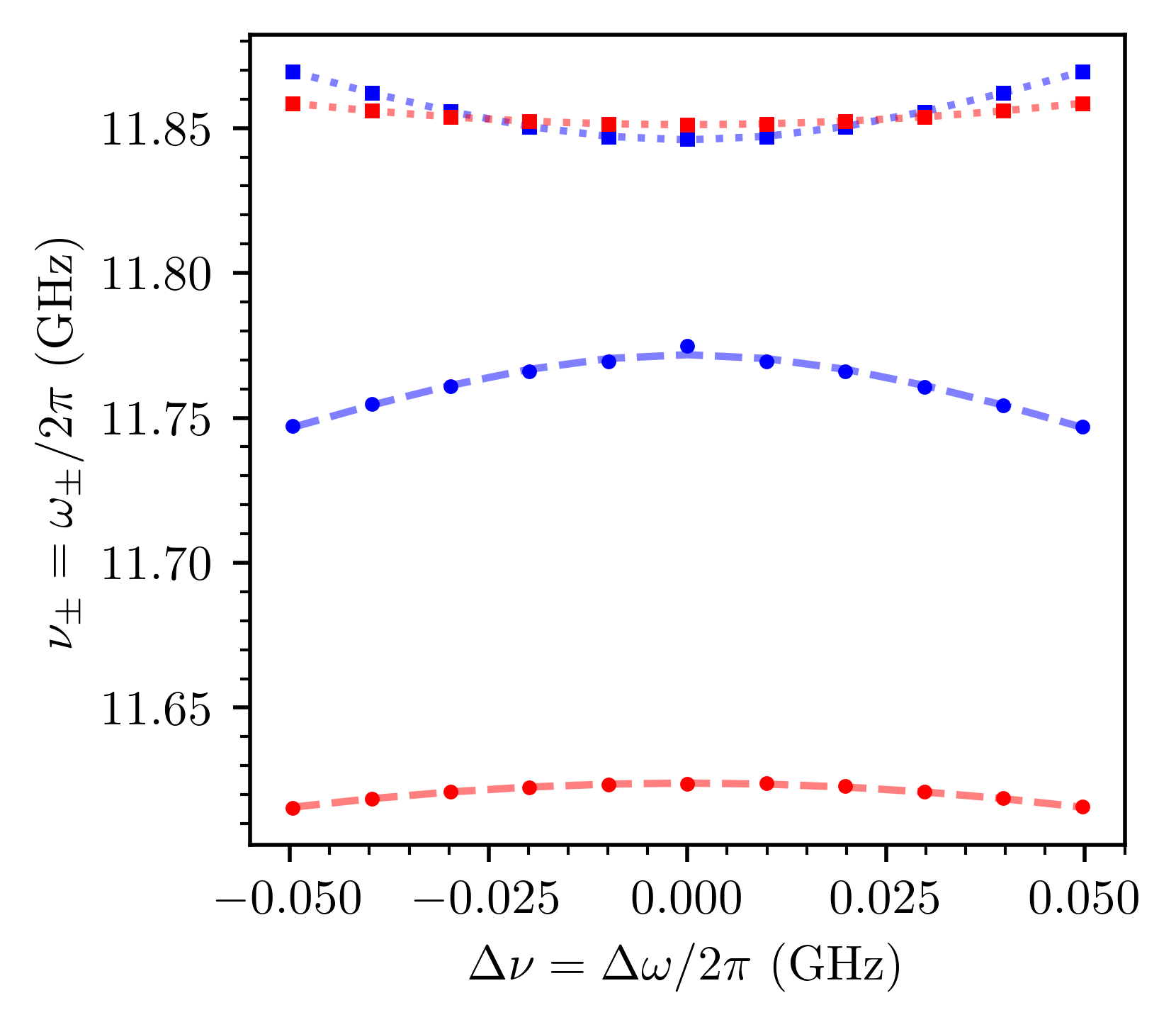}
    \caption{Resonant frequency of the symmetric $\nu_+ = \omega_+ / 2\pi$ (circular markers) and antisymmetric $\nu_- = \omega_- / 2\pi$ (square markers) modes of two strongly coupled 2-cell cavities verses detuning $\Delta \nu = \Delta \omega / 2\pi$. Data in blue corresponds to a cavity with $\ell = 38.75$ mm, while data in red corresponds to a cavity with $\ell = 38.5$ mm. Both cavities had $R=19.5$ mm. Detuning was accomplished by varying the radius $r$ of the cavity's two tuning rods in an antisymmetric fashion. We matched the change in tuning rod radius $\delta r$ to detuning $\Delta \nu = \Delta \omega / 2\pi$ by interpolating over simulations of several single-cell cavities with different tuning rod radii. The dashed (dotted) lines represent a fit to Equation \ref{eq:mode_hybridizations} in its symmetric (antisymmetric) form. For the $\ell = 38.75$ mm cavity, the symmetric mode has a form factor of 0.79 with no detuning and 0.64 with maximum detuning. For the $\ell = 38.5$ mm cavity that has a strong coupling, the symmetric mode has a form factor of 0.69 with no detuning and 0.67 with maximum detuning. With no coupling the form factor drops below 0.4 with a sight ($\sim$MHz) detuning.}
    \label{fig:mode_splitting}
\end{figure}

The simulated resonant frequencies $\omega_+$ and $\omega_-$ for our 2-cell model should vary with $\Delta \omega$ according to Equation \ref{eq:mode_hybridizations}. Since our simulations adjust $\Delta \omega$ via asymmetric variation of the rod radii, we first establish the relationship between $\delta r$ and $\Delta \omega$. To do this, we performed an auxiliary study tracking $\omega_+$ for a single-cell cavity as we varied the tuning rod radius $r$. Figure \ref{fig:radius_freq_relationship} shows the results and a linear fit.
Using this linear relationship between $\delta r$ and detuning $\Delta\omega$, we then plotted $\omega_+$ and $\omega_-$ for our two-cell system versus $\Delta \omega$, as shown in Figure \ref{fig:mode_splitting}. Blue data points correspond to a simulated cavity with $\ell = 38.75$ mm while red data points correspond to a similar cavity with $\ell = 38.5$ mm. Circular markers denote $\omega_+$, and square markers denote $\omega_-$. 
The dashed (dotted) curves under each set of data points correspond to a fit to Equation \ref{eq:mode_hybridizations} in its symmetric (antisymmetric) form. Table \ref{table:model_fit_parameters} summarizes the model fit parameters and uses them to compute the splitting $\Omega_{fit}$ in the $\omega_0 \gg \omega_c$ limit. It also compares this result to the observed splitting measured directly from the simulation data $\Omega_{measure}$. (Note that values simulated with COMSOL-RF are returned as absolute frequencies and thus differ from the angular frequencies predicted by our model by a factor of $2\pi$. This is noted where appropriate in the figures and table.)

\begin{table*}[]
\caption{Parameters extracted from fitting the data in Figure \ref{fig:mode_splitting} to Equation \ref{eq:mode_hybridizations}.}
\begin{tabular}{|c|c|c|c|c|}
\hline
Cell Separation $\ell$ (mm) & $\omega_c/2\pi$ (GHz) & $\omega_0/2\pi$ (GHz) & $\Omega_{fit}/2\pi \sim \frac{\omega_c^2}{2\pi\omega_0}$ (MHz) & $\Omega_{measure}/2\pi$ (MHz) \\ \hline
38.75   & 0.917                     & 11.772                    & 71.388                                                         & 71.385                 \\ \hline
38.5                                 & 1.804                     & 11.537                    & 279.915                                                        & 227.391                \\ \hline
\end{tabular}
\label{table:model_fit_parameters}
\end{table*}

Notice that, as expected from our theoretical calculations, $\omega_+$ is always lower in frequency than the $\omega_-$. Furthermore, we see that the mode splitting $\Omega$ increases significantly as we increase the coupling (decrease $\ell$). Thus, as predicted, we find that cavities with strong coupling exhibit an increased robustness to detuning of their individual modes. 

The two modes never cross, demonstrating the avoided crossing phenomenon \cite{novotny10} mentioned above. In fact, as the detuning continues to increase, the two modes diverge further until $\Delta\omega \sim \Omega$. At this point, the form factor reduces significantly as the modes cease to combine coherently. 

\section{The Beehive Geometry}

\begin{figure}
\includegraphics[width=0.49\textwidth,left]{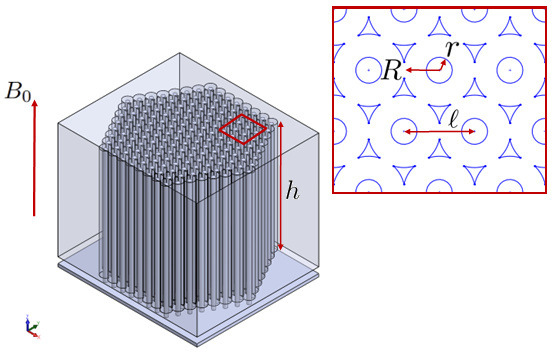}
\caption{\label{fig:concept} A conceptual 3D rendering of a 169-cell beehive haloscope. Each cell consists of a conventional coaxial cavity, whose parallel axial directions can be aligned with a single magnetic field $B_0$ during a typical axion search. Each cell overlaps with its neighbors, allowing the entire system to behave as a single resonator. Rods placed inside each cell can be moved in the $xy$-plane to tune the resonant frequency of the system. Notice that the rods are attached to a single plate, allowing them to be moved in unison during tuning. We denote the cell radius with $R$, the tuning rod radius with $r$, the cell separation with $\ell$, and the height of the cavity with $h$.}
\end{figure}

In Figure \ref{fig:concept}, we introduce the geometry of a beehive haloscope. The system consists of an array of parallel coaxial cavities, whose axial directions can be simultaneously aligned with a magnetic field $\boldsymbol{B_0}$. However, unlike other multi-cavity proposals and experiments, the beehive haloscope is typified by an overlapping of neighboring cavities. The overlap creates small passages between adjacent cavities that permit interactions between their individual resonant modes. As we will show via 2D and 3D FEA below, the individual interacting modes found in each ``cell" are strongly coupled, which results in the formation of a single TM$_{010}$ mode for the entire system. We can think of the beehive geometry as a hexagonal lattice of strongly coupled oscillators behaving according to the mechanics we explored in section IB. 

Within each cell is a tuning rod which can be moved in the $x$- and $y$-directions to change the resonant frequency of the shared TM$_{010}$ mode. We envision the rods attached to a single platform so that they can be moved in unison. The rods should be precisely centered within each cell before tuning begins. Idealized tuning involves moving the shared platform so that the rods are identically positioned within their respective cells. We will explore deviations from this perfect positioning assumption as a part of our 2D simulations. In 3D, form factor degradation due to the necessary gap between tuning rods and the endcap of the cavity warrants investigation. We will also explore the effects of minor errors in the angle of the rods.

The beehive cavity has several free geometric parameters (see Figure \ref{fig:concept}). These include the radius of each cell $R$, the separation between cells $\ell$, the rod radius $r$, the height of the cavity $h$, and the number of cells $n$. Our scheme presupposes $\ell<2R$ so that adjacent cells overlap.  

\section{Computational Analysis}

\subsection{2D Simulations: Scalability, Tunability, and Resilience}

\subsubsection{2D Simulation Principles}

We begin our analysis of beehive haloscopes with a series of 2D FEA. These simulations allow us to explore the scalability, tunability, and resilience of the experiment before moving to more computationally-intensive 3D simulations (see Section IV below). All results were generated using COMSOL-RF on a workstation with 18 cores and 160 GB of available RAM. While the beehive haloscope geometry under consideration is specific to each simulation set, we note here that all simulations were conducted using COMSOL-RF's Cu material (with the surface conductivity increased to $3\times10^8$ S/m to simulate behavior at cryogenic temperatures) for the boundaries of the cavity and tuning rods and Air material for the internal spaces of the cavity.

Throughout this section, we will regularly explore the impact of different geometric configurations on the cavity's unloaded quality factor $Q_0$ and form factor with respect to the shared TM$_{010}$ mode of the system $C_{010}$. To use these quantities in 2D, we modify their typical form as follows. We define the unloaded quality factor as
$Q_0$ as
\begin{align}
    Q_0 &= \frac{\textrm{Energy Stored in Cavity}}{\textrm{Energy Dissipated at Cavity Wall}} \nonumber \\
    Q_0 &= \frac{\int_A|E_z|^2dA}{\int_\ell|E_t|^2d\ell},
\end{align}
where $E_z$ is the $z$-component of the electric field and $E_t$ is the component of the electric field tangent to the cavity wall. In the numerator we have replaced the volume integral appropriate in 3D with an area integral over the active area of the cavity. Similarly, in the denominator, we have replaced the area integral found in the 3D case with a line integral. These integral substitutions are also present in our working form factor definition
\begin{align}
    C_{010} = \frac{(\int_A E_z dA)^2}{A\int_AE^2dA}
\end{align}
(compare to Equation \ref{eq:C_Definition}).

\subsubsection{Scalability}

\begin{figure*}
    \centering
    \includegraphics[width=0.6\textwidth]{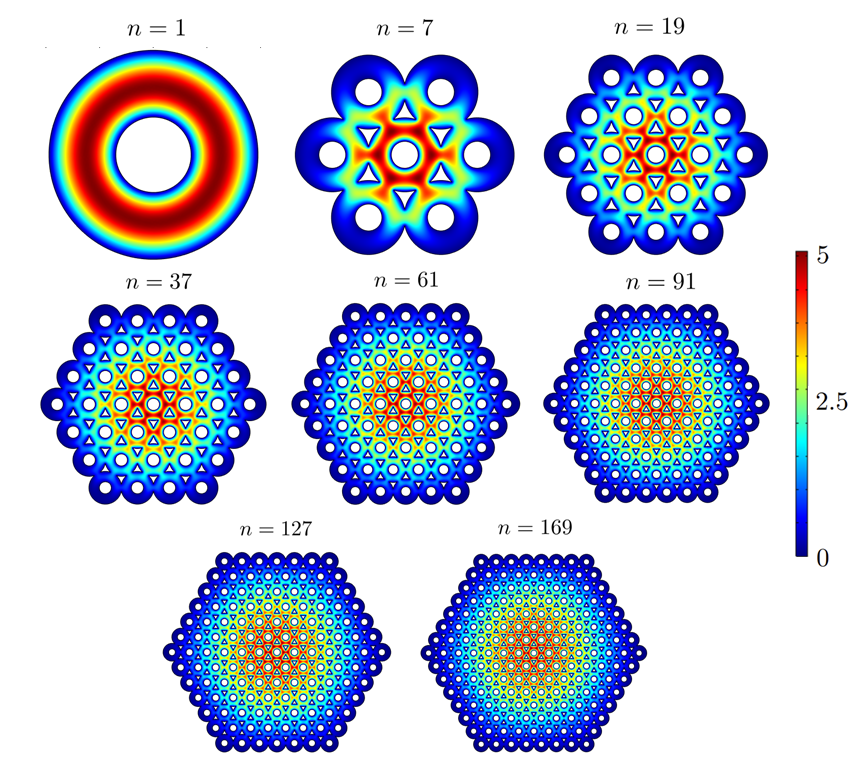}
    \caption{Map of $E_z$ for the $C_{010}$ mode of beehive cavities with $n$ = 1, 7, 19, 37, 61, 91, 127, and 169 cells. In each case, $\ell = 38$ mm, $R = 19.5$ mm, $r = 7$ mm.}
    \label{fig:scalability_efields}
\end{figure*}

One of the primary advantages of the beehive halosocope is the ability to arbitrarily increase the volume by increasing the number of coupled cells, a characteristic we refer to as the \textit{scalability} of the cavity. To be truly scalable, beehive should not see a significant decrease in $Q_0$ and $C_{010}$ as the number of cells increases. In Figure \ref{fig:scalability_efields}, we explore the effect of increasing the number of cell-rings in the system, beginning with a single cell (\textit{i.e.}, a typical single-cavity haloscope) and continuing through a system with seven additional rings.
The largest system we studied has a total of 169 cells. The cavity geometry in all cases was $R=19.5$ mm, $r = 7$ mm, and $\ell = 38$ mm. If we assume that our 2D system is a slice of a 1 m tall cavity, $n$ corresponds almost exactly to the volume of the system in liters for this particular geometry. 

\begin{figure}
    \includegraphics[width=0.5\textwidth]{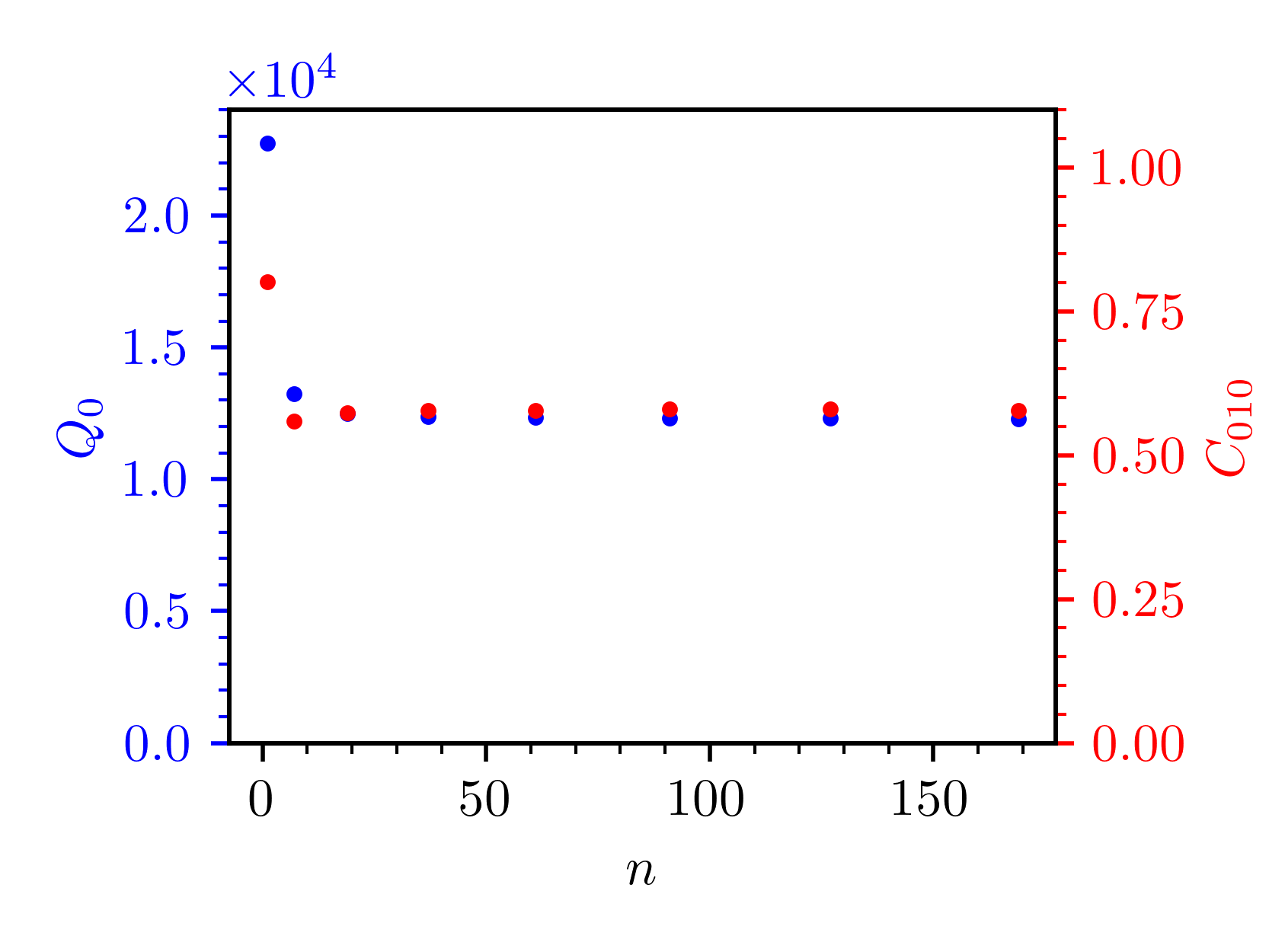}
    \caption{Unloaded quality factor $Q_0$ and form factor $C_{010}$ vs. number of cells $n$ for beehives with $\ell = 38$ mm, $R = 19.5$ mm, and $r = 7$ mm.}
    \label{fig:scalability_results}
\end{figure}

Figure \ref{fig:scalability_results} summarizes the computed form and quality factors for the simulations shown in Figure \ref{fig:scalability_efields}.
Notice that moving from a single-cavity haloscope ($n = 1$) to a beehive haloscope ($n > 1$) leads to a reduction in both parameters. However, after this initial drop (to still-respectable values), these two key parameters remain relatively stable, all while the volume of the system increases dramatically.

\subsubsection{Tunability}

\begin{figure}
    \centering
    \includegraphics[width=0.305\textwidth]{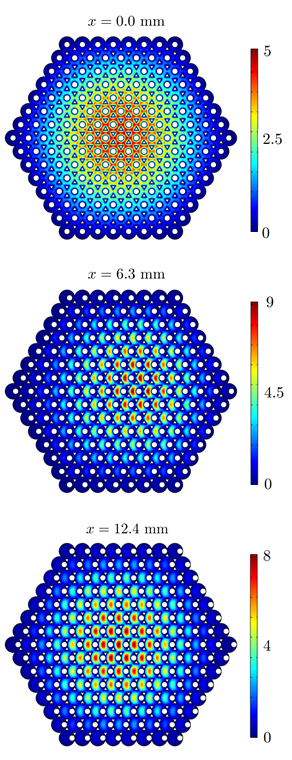}
    \caption{Map of $E_z$ for the $C_{010}$ mode of an $n = 169$-cell cavity at tuning points $x = 0$ mm, $x = 6.3$ mm, and $x = 12.4$ mm. In each case, $\ell = 38$ mm, $R = 19.5$ mm, and $r = 7$ mm.}
    \label{fig:tuning_E_fields}
\end{figure}

\begin{figure}
    \includegraphics[left]{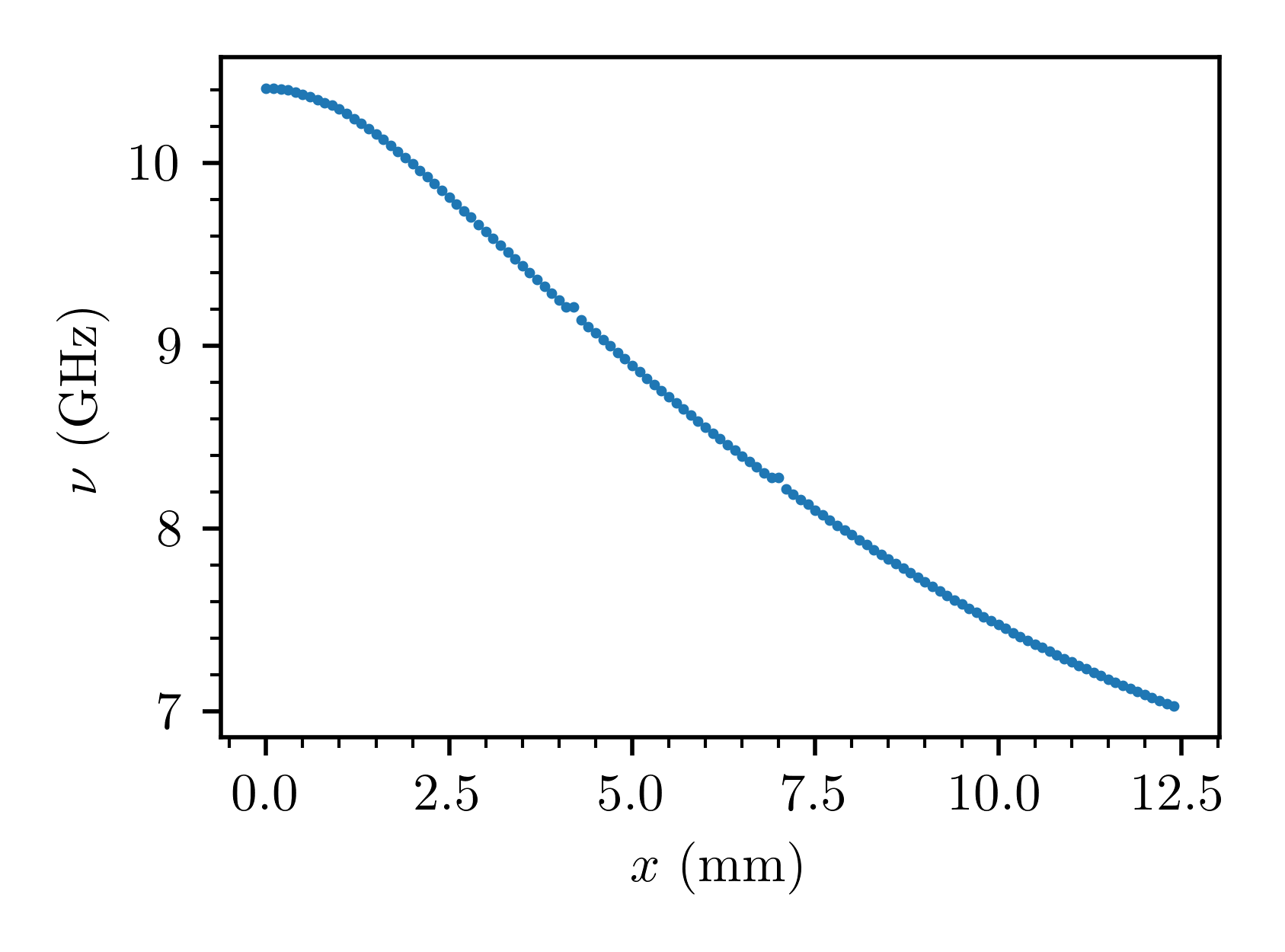}
    \caption{Resonant frequency $\nu$ vs. horizontal rod displacement $x$ for an $n=$169-cell cavity with $\ell = 38$ mm, $R=19.5$ mm, and $r = 7$ mm, assuming perfect rod positioning.}
    \label{fig:ideal_tuning_freq}
\end{figure}

\begin{figure}
    \includegraphics[left]{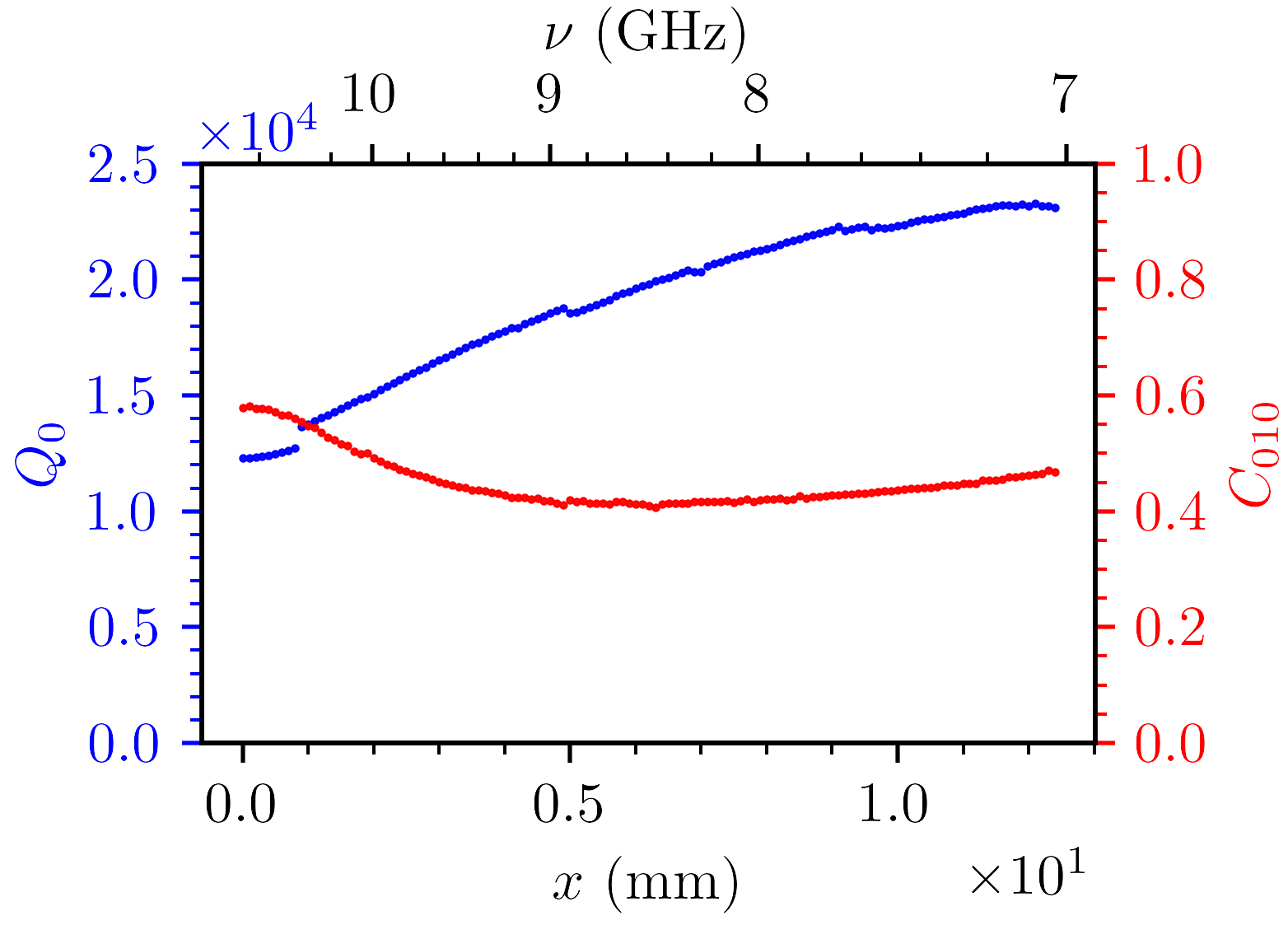}
    \caption{Unloaded quality factor $Q_0$ and form factor $C_{010}$ vs. horizontal rod displacement $x$ for the same cavity as described in Figure \ref{fig:ideal_tuning_freq}.}
    \label{fig:ideal_tuning_Q0_and_C}
\end{figure}

We now explore the \textit{tunability} of a beehive halsocope, again using a geometry of $R = 19.5$ mm, $r = 7$ mm, and $\ell = 38$ mm. For this study, we fixed $n$ at $169$. As a reminder, we define ideal tuning of the cavity as the case in which each tuning rod occupies the exact same $x$- and $y$- position relative to its cell center. Figure \ref{fig:tuning_E_fields} depicts $E_z$ throughout the cavity at several different tuning positions.
In Figure \ref{fig:ideal_tuning_freq}, we plot the resonant frequency $\nu$ of the TM$_{010}$ mode vs. ideal horizontal displacement of the tuning rods $x$. Figure \ref{fig:ideal_tuning_Q0_and_C} summarizes the values of $Q_0$ and $C_{010}$ over the same tuning range.
From these results, we observe the tuning range of this particular geometry, roughly 7-10.5 GHz. This frequency range corresponds to a mass range of 28.9-43.4 $\mu$eV and was chosen intentionally to avoid parameter space that is currently under exploration by actively scanning experiments.

We see that the quality factor increases as we move the rods horizontally and progress to lower frequencies ($\sim 12,000 \rightarrow~\sim 23,000$). This increase in quality factor arises because the high-electric field region of the mode makes contact with a smaller portion of the cavity/rod walls when the rods are displaced (lower frequencies). Thus, there is less dissipation. This behavior can be clearly seen in Figure \ref{fig:tuning_E_fields}. The form factor begins at $\sim0.575$, falls to a minimum of $\sim0.4$ at half the displacement distance, before rising again to around $\sim0.475$. 

We also performed simulations in which we moved the rods in the $y$-direction. In this case, we found similar quality and form factors and accessed no additional frequency space. 

\begin{figure}
    \includegraphics[left]{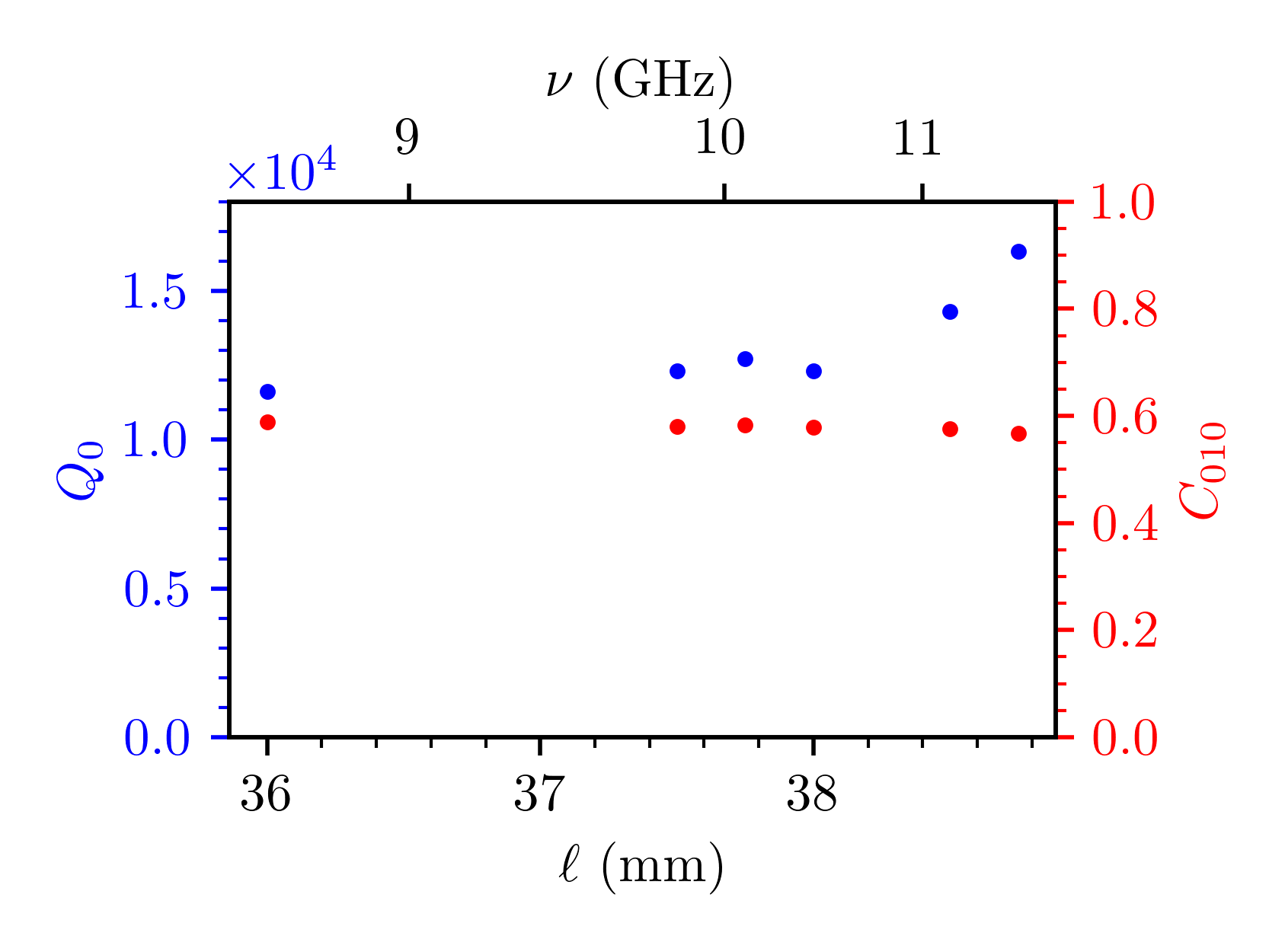}
    \caption{Unloaded quality factor $Q_0$ and form factor $C_{010}$ vs. cell separation $\ell$. An additional horizontal axis shows the initial resonant frequency of the simulated cavities. Cavity dimensions were $n= 169$, $R = 19.5$ mm, $r = 7$ mm, and $n = 169$.}
    \label{fig:tunability_changing_coupling}
\end{figure}

We also note that the coupling between the cells, which can be adjusted during cavity manufacturing by changing the distance $\ell$ between adjacent cells, also affects the starting frequency, quality factor, and form factor of the system. Furthermore, as we discussed in Section IB, increasing the coupling also increases the frequency splitting between the primary symmetric and antisymmetric modes of the cavity. Since the width of this mode split functions as the effective tolerance criterion for guaranteeing in-phase oscillations of the cells, it is a natural system parameter to explore in more detail. (Note that we investigate this phenomenon further in Section IIIA4.) In Figure \ref{fig:tunability_changing_coupling}, we plot $Q_0$ and $C_{010}$ vs. $\ell$ for cavities with $n = 169$, $R = 19.5$ mm, and $r = 7$ mm.
We include the simulated resonant frequency of the TM$_{010}$ mode as an additional horizontal axis. Notice that increasing the coupling (\textit{i.e.}, decreasing $\ell$) leads to a small drop in frequency and quality factor. The form factor remains stable across the simulation set.

At this point, we comment briefly on the geometric parameters available to the experimentalist for setting the initial frequency of the cavity (\textit{i.e.}, resonant frequency when rods are centered in their cells). The cell radius $R$, rod radius $r$, and cell separation $\ell$ all ultimately impact the system's resonant frequency and quality factor. Additionally, we will see below that $\ell$ also affects the ability of the cavity to resist drops in form factor due to non-ideal tuning rod positioning. Therefore, it will be necessary to carefully consider the optimal combination of these three parameters in light of potential manufacturing and positioning errors when constructing a beehive cavity to cover a particular frequency range.  

\subsubsection{Resilience}

As suggested above, construction of a real-world experiment will necessarily introduce imperfections in the geometry of the beehive haloscope. Common imperfections include those introduced during the machining process, as well as others which arise due to irregularities in the assembly of the cavity and its tuning apparatus. Here we use our 2D simulation technique to explore two critical imperfections: (1) variations in tuning rod radii within a machining tolerance and (2) variations in tuning rod position within an assembly tolerance. Both cases may result in detuning of the modes supported by the cavity cells, eventually leading to a breakdown in coherence. We seek to explore the behavior of the system at coherence breakdown and define the minimum tolerance level required to maintain the form factor and quality factor at 20\% of their levels in an ideal cavity. Furthermore, our analytical model for two coupled cells (see Section IB) suggests that coherence is maintained as long as detuning remains under the order of the frequency splitting $\Omega$ between the symmetric and antisymmetric modes of the system when no detuning is present. This is opposed to the behavior of uncoupled cavities, which lose coherence when detuning exceeds the cavity linewidth. Here, we investigate the applicability of this model to the $n\gg 2$ case. Included is an analysis of the effect of cavity coupling on the required tolerance. In all, these studies aid in quantifying the \textit{resilience} of the beehive haloscope scheme and will help to inform our selection of a final geometry for the experiment.

To study the impact of imperfections in tuning rod radius, we stochastically select the $n$ tuning rod radii needed to populate a cavity from a truncated normal distribution with probability density function
\begin{align*}
    P(r) = \frac{1}{\delta r_{max}/2} \frac{\frac{1}{\sqrt{2 \pi}} \exp(-\frac{1}{2} \frac{r-r_{ideal}}{\delta r_{max}/2})}{\Phi(-2)-\Phi(+2)}.
\end{align*}
Here, $r$ is the radius of the rod, $\delta r_{max}$ is the maximum permissible deviation in rod radius (\textit{i.e.}, the tolerance), and $\Phi(x) = \frac{1}{2} ( 1 + \textrm{erf} (x / \sqrt{2}) )$ is the cumulative distribution function of the normal distribution. This distribution has its mean at $r_{ideal}$, a standard deviation of $\delta r_{max}/2$, and limits $[r-\delta r_{max}, r+\delta r_{max}]$. This approach assumes that imperfections in the rod radii are independent random variables, with a distribution truncated by a tolerance verification procedure.

\begin{figure}
    \centering
    \includegraphics[width=0.48\textwidth]{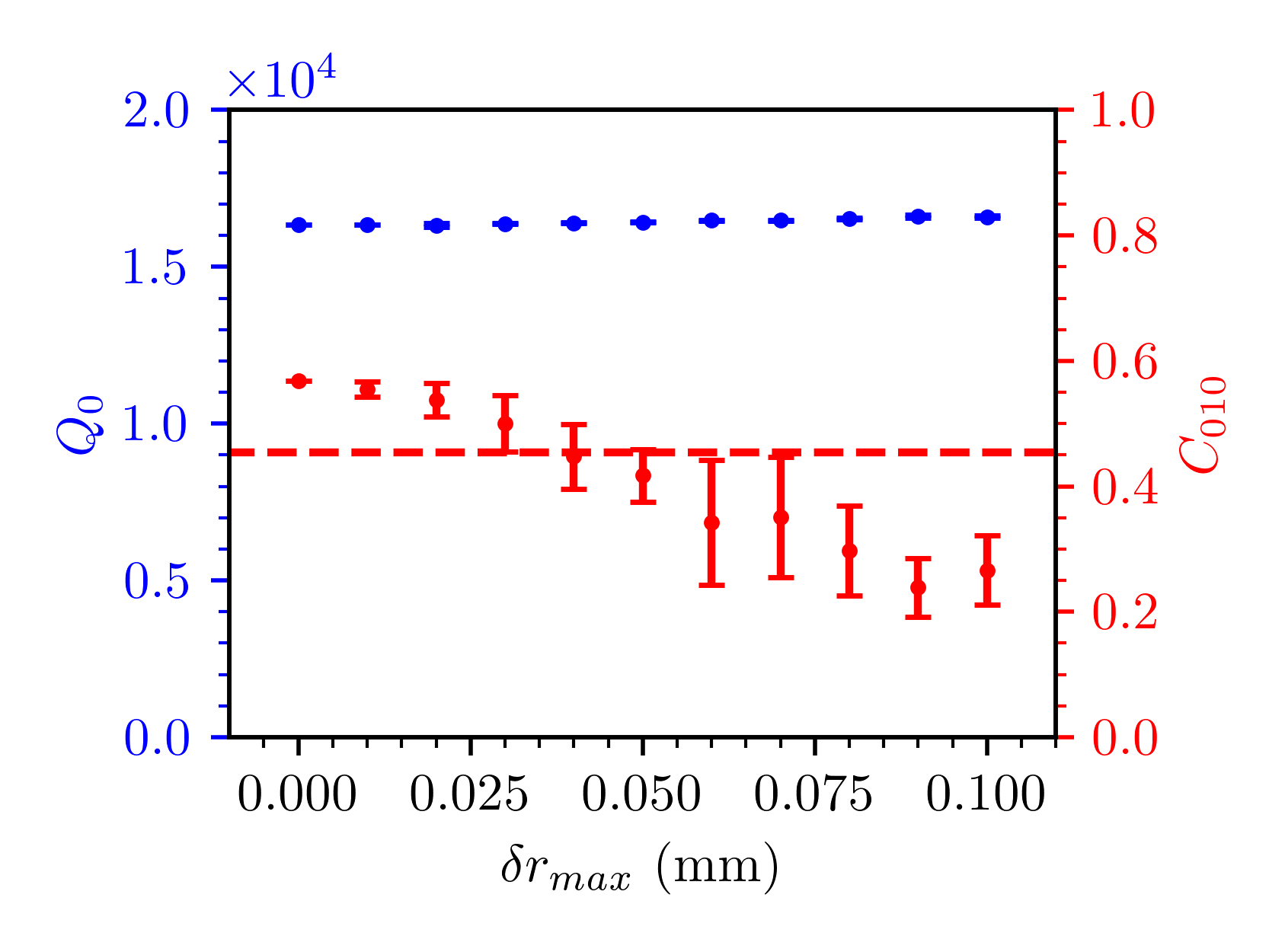}
    \caption{Unloaded quality factor $Q_0$ and form factor $C_{010}$ vs. maximum rod radius variation $\delta r_{max}$. Cavity dimensions were $n=169$, $\ell = 38.75$ mm, $R=19.5$ mm, $r_{ideal} = 7$ mm. The dashed red line indicates a 20\% reduction in form factor compared to the simulated form factor observed in an ideal cavity with the same dimensions.}
    \label{fig:random_radii}
\end{figure}

We then fed the $n$ radii into a custom Python script, which uses the \texttt{ezdxf} package to quickly generate 2D CAD drawings of the haloscope. These CAD drawings were simulated in COMSOL-RF using our typical setup. In Figure \ref{fig:random_radii}, we show the effect of varying $\delta r_{max}$ on the unloaded quality factor $Q_0$ and form factor $C_{010}$ of an $n=169$-cell cavity with $\ell=38.75$ mm, $R=19.5$ mm, and $r_{ideal} = 7$ mm.
Each data point represents the appropriate quantity mean for the highest form factor mode across 10 different simulated cavities. The error bars give the corresponding standard deviations. The unloaded quality factor remains stable across the $\delta r_{max}$ range shown. This indicates that variations in the machining tolerance of the rods should have no appreciable impact on quality factor up to the 100 $\mu$m level. The form factor, however, decreases as $\delta r_{max}$ increases. In the figure, we include a dashed red line to indicate 20\% reduction in form factor with respect to a cavity with no rod radius variation. By performing a simple linear interpolation between the simulated data points, we expect 20\% form factor reduction to occur at $\delta r_{max} \approx 0.039$ mm.

\begin{figure}
    \centering
    \includegraphics[width=0.48\textwidth]{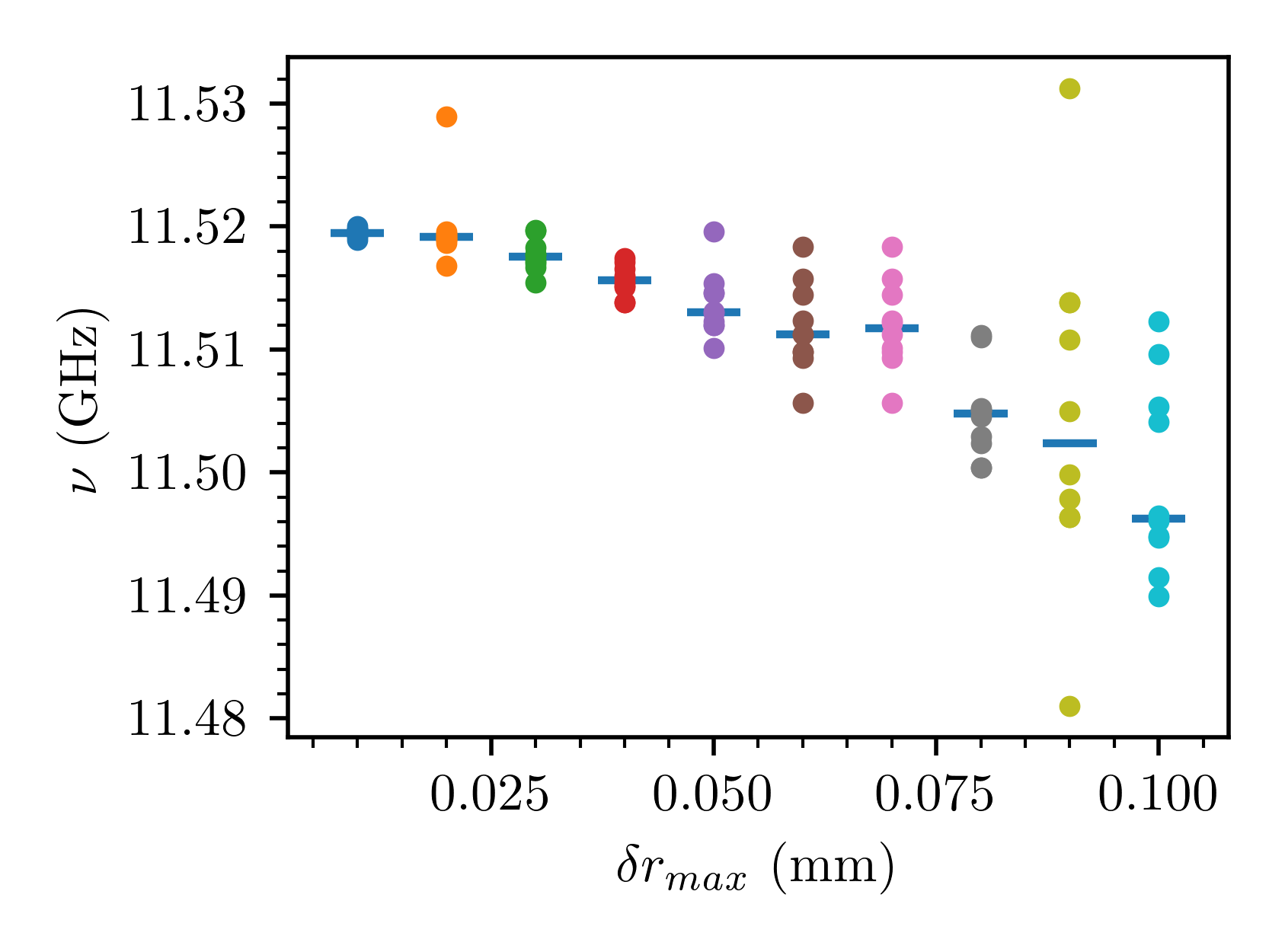}
    \caption{Simulated resonant frequencies $\nu$ for the 10 beehive cavities simulated at each maximum rod radius variation $\delta r_{max}$. Cavity dimensions were $n=169$, $\ell = 38.75$ mm, $R = 19.5$ mm, $r_{ideal} = 7$ mm. The small horizontal lines indicate the median frequency for each set. Notice the increase in variability as $\delta r_{max}$ increases.}
    \label{fig:radius_variation_frequencies}
\end{figure}

We note that introducing rod radius variance also produces variation in the frequency of the highest form factor mode of the cavity. In Figure \ref{fig:radius_variation_frequencies}, we plot the simulated eigenfrequency of the highest form factor mode from each of the 10 cavities generated at each $\delta r_{max}$ value. We include blue horizontal lines to indicate the median.
The spread of simulated frequencies increases significantly (up to the 10s of MHz level) as $\delta r_{max}$ grows. 

To gain more insight into why the form factor decreases and frequency variation increases, we sought to simulate the response of the beehive cavity to an axion signal. Following the approach of \cite{dmradiocollaboration2023electromagnetic} and \cite{Jeong:2023bqb}, we used COMSOL-RF's \textit{External Current Density} node to introduce an axion field-induced current density
\begin{align*}
    \boldsymbol{J_e} \approx i g_{a \gamma \gamma} a_0 \omega_a \frac{\boldsymbol{B_0}}{Z_0}.
\end{align*}
In this equation, $g_{a \gamma \gamma}$ is the coupling for axion-photon interactions, $a_0$ is the amplitude of the axion field, $\omega_a$ is the angular frequency of the axion field, $\boldsymbol{B_0}$ is the external magnetic field, and $Z_0 = \sqrt{\mu_0 / \varepsilon_0}$ is the impedance of free space. With the introduction of this external current density, our simulations effectively report the response of the cavity to the electric and magnetic fields $\boldsymbol{E_r}$ and $\boldsymbol{B_r}$ reacted by dark matter axions in the presence of a constant magnetic field $\boldsymbol{B_0} = \boldsymbol{B_z}$. These fields are given by the analytic approximation 
\begin{align*}
    \varepsilon \nabla \cdot \boldsymbol{E_r} &= \frac{g_{a \gamma \gamma}}{Z_0} (\nabla a) \cdot \boldsymbol{B_0} \approx 0, \\
    \nabla \cdot \boldsymbol{B_r} &= 0, \\
    \nabla \times \boldsymbol{E_r} + \dot{\boldsymbol{B_r}} &= 0, \\
    \frac{1}{\mu} \nabla \times \boldsymbol{B_r} - \varepsilon \dot{\boldsymbol{E_r}} &= -\frac{g_{a \gamma \gamma}}{Z_0} (\dot{a} \boldsymbol{B_0} + (\nabla a) \times \boldsymbol{E_0}) \\
    &\approx \frac{i \omega_a}{Z_0} (g_{a \gamma \gamma} a_0 e^{-i \omega_a t}) \boldsymbol{B_0}.
\end{align*}
Under these conditions, the axion conversion power $P_{a \gamma \gamma}$ can be computed using
\begin{align*}
    P_{a \gamma \gamma} = P_{ohmic} = \frac{1}{2} \textrm{Re} \bigg{[} \int \boldsymbol{J_a^*} \cdot \boldsymbol{E_r} dA \bigg{]},
\end{align*}
where we assume the axion-induced current density $\boldsymbol{J_a} = \boldsymbol{J_e}$ \cite{KIM2019100362}. Note that we have again converted the reported volume integral into an area integral for use in 2D simulations. The resulting units are therefore in W/m. Multiplying by an assumed cavity height returns a true power in W.

\begin{figure*}
    \centering
    \includegraphics[width=0.7\textwidth]{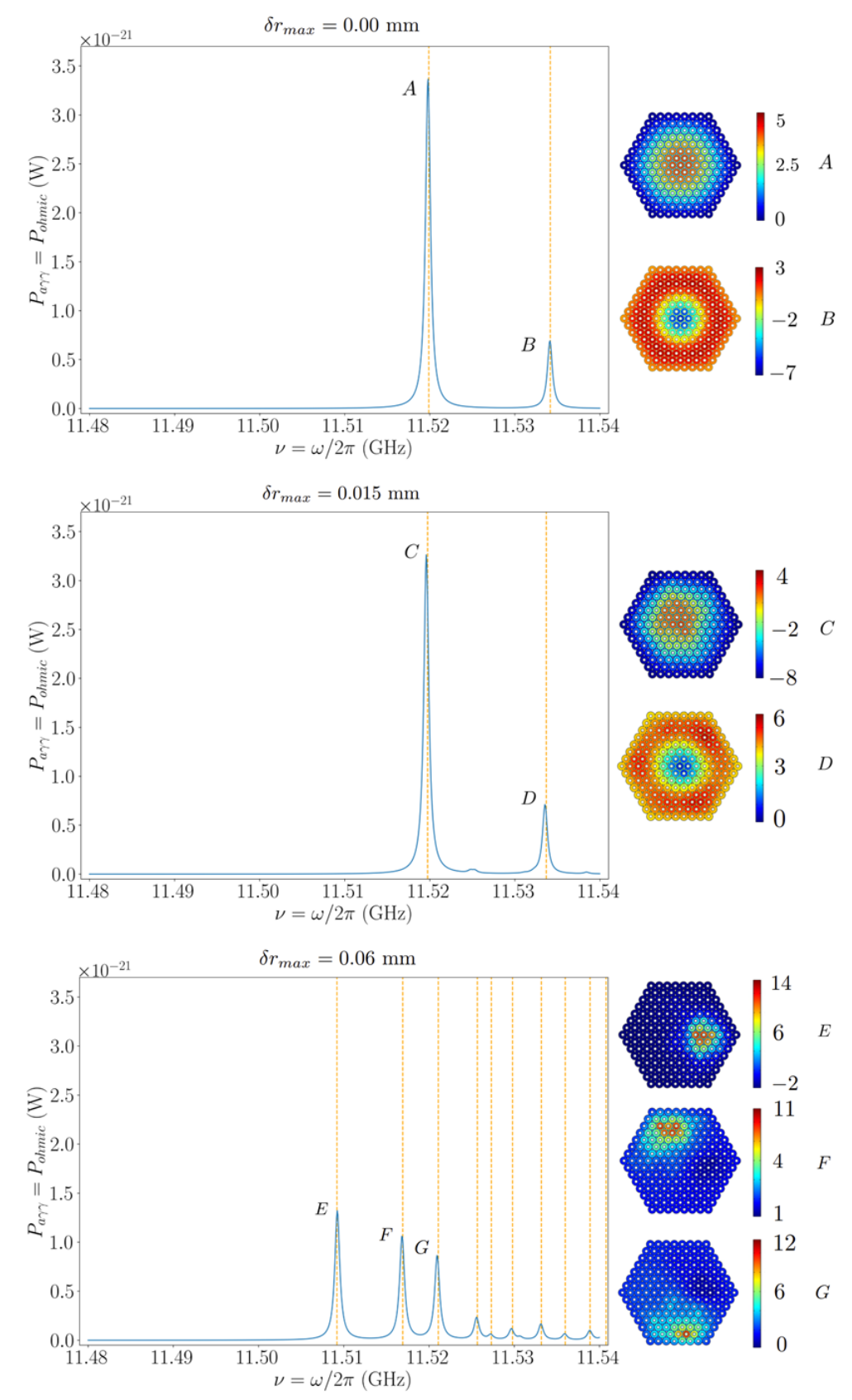}
    \caption{Axion conversion power $P_{a \gamma \gamma}$ simulated by introducing an external current density $\boldsymbol{J_e}$ to COMSOL-RF's \textit{Frequency Domain} solver. All three simulated cavities had $n=169$, $\ell = 37.75$ mm, $R = 19.5$ mm, and $r_{ideal} = 7$ mm. To report a power in W from 2D simulations, we assumed that the simulated cavity represents a slice of a 1 m tall 3D haloscope. The plot labeled $\delta r_{max} = 0.00$ mm corresponds to a cavity with no rod radius variation. The plot with $\delta r_{max} = 0.015$ mm corresponds to a cavity with radius variations that are within the machining tolerance window but produce frequency variations much larger than the cavity linewidth. The plot with $\delta r_{max} = 0.06$ mm is an example of a cavity with radius variations larger than the optimal machining tolerance. The orange-colored vertical dashed lines indicate the resonant frequencies of all modes found using COMSOL-RF's \textit{Eigenfrequency} solver with a form factor of 0.01 or greater. The $z$-component of the electric field for select modes (also found using the \textit{Eigenfrequency} solver and labeled with letters) are plotted to the right.}
    \label{fig:external_axion_response}
\end{figure*}

In Figure \ref{fig:external_axion_response}, we plot (solid blue curves) the simulated conversion power returned by COMSOL's \textit{Frequency Domain} solver for an ideal cavity ($\delta r_{max} = 0.00$ mm) and cavities with $\delta r_{max} = 0.015$ mm and $\delta r_{max} = 0.06$ mm.
In all three cases, the cavity geometry was $n = 169$, $\ell = 38.75$ mm, $R = 19.5$ mm, and $r_{ideal} = 7$ mm, and the strength of the $\hat{z}$-aligned magnetic field was 10 T. We also assumed KSVZ axions with model-dependent coefficient $g_{\gamma} = 0.97$. To achieve a result in W, we assume that our 2D simulation represents a slice of a 3D cavity with height $h = 1$ m. The orange-colored vertical dashed lines indicate the resonant frequencies of all modes found using COMSOL's \textit{Eigenfrequency} solver with a form factor of 0.01 or greater. Plots of $E_z$ found with the \textit{Eigenfrequency} solver are plotted to the right for select modes.

In the ideal cavity case, we see two modes, labeled $A$ and $B$. $A$ represents the fundamental TM$_{010}$ mode of the cavity, as indicated by the electric field plot to the right of the graph. $B$ represents a higher order mode that also has appreciable alignment with the magnetic field. When we increased the rod radius variation to $\delta r_{max} = 0.015$ mm---a value that is within the machining tolerance window suggested by 20\% form factor reduction but produces frequency detuning much larger than the cavity linewidth---the mode structure remains nearly the same (see modes labeled $C$ and $D$). In the cavity with $\delta r_{max} = 0.06$ mm, where we have moved beyond the machining tolerance window, the mode landscape becomes more complicated. Modes $E$, $F$, and $G$ all have form factors above 0.1. Looking at the electric field plots to the right, we see that they represent coherence between small selections of cells. Since detuning between individual cells is high, the cavity is unable to resonate as a single unit. Instead, small groups of adjacent cells, whose resonant frequencies are, by chance, within the coherence limit, form reduced-form factor modes. This behavior explains why we see both a reduction in form factor and increase in variation in the resonant frequency of the highest form factor mode as $\delta r_{max}$ increases. For form factor, the original high-$C$ mode has been divided into several low-$C$ modes. For frequency, the random generation of rod radii produces different groups of coherent cells, each with slightly different resonant frequencies, for each haloscope model.

In Section IB, we reported the simulation-derived relationship between detuning $\delta \omega$ and rod radius variation $\delta r$ (see Figure \ref{fig:radius_freq_relationship}). Extrapolating from this relationship, the rod radius variation which would produce a detuning on the order of the frequency splitting $\Omega /2\pi = 71.385$ MHz 
(see Table \ref{table:model_fit_parameters})
of a $\ell = 38.75$ mm 2-cell cavity is $\delta r \approx 0.23$ mm. As mentioned above, our simulations in the 169-cell case show that decoherence begins to occur at (using the 20\% form factor reduction point) $\delta r_{max} =$ 0.039 mm. This represents roughly an order of magnitude decrease in the required tolerance vis-\`a-vis the model predictions. This reduction is likely due to uncaptured behavior introduced by the increased number of cell couplings (\textit{e.g.}, interior cells are coupled to six neighbors, rather than one as in the two cell model). However, the 169-cell cavity's decoherence point is much larger than the $\delta r_{max} = 5.1\times 10^{-4}$ mm decoherence point suggested by a representative cavity linewidth (assuming, for example, $Q_0 \sim 16,000$ and $\omega_0/2\pi\sim 8$ GHz).

\begin{figure}
    \centering
    \includegraphics[width=0.48\textwidth]{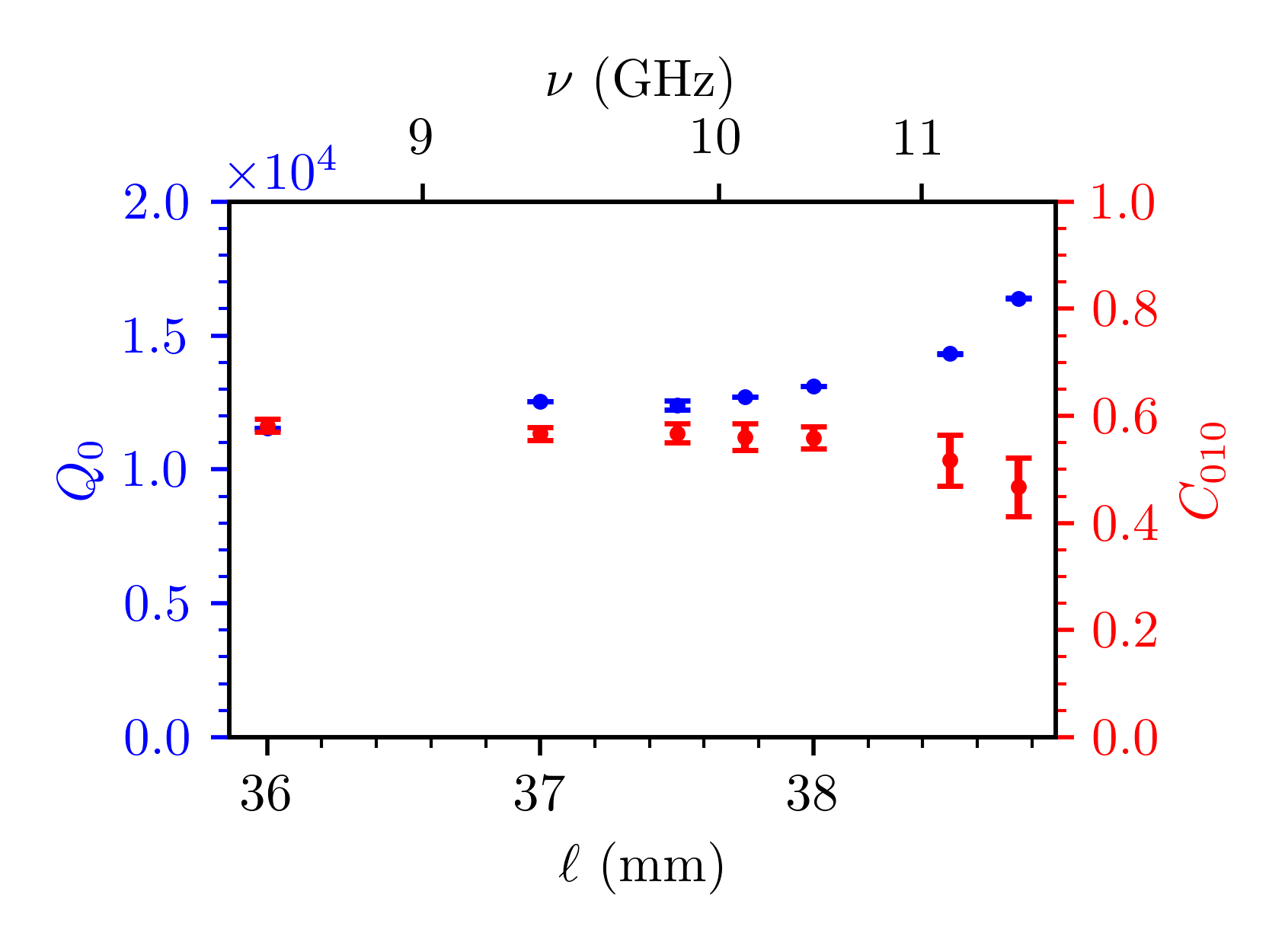}
    \caption{Unloaded quality factor $Q_0$ and form factor $C_{010}$ vs. cell separation $\ell$ for cavities with 0.039 mm of rod radius variation. This displacement error results in a 20\% form factor reduction for a cavity with $\ell = 38.75$ mm. Other cavity dimensions were $n=169$, $R = 19.5$ mm, and $r = 7$ mm.}
    \label{fig:radii_variance_20_percent_reduction}
\end{figure}

Our model also predicted that increasing the coupling between cells (decreasing $\ell$) would increase the frequency splitting and, by extension, relax the required rod radius tolerance. To ascertain whether the 169-cell beehive follows this behavior qualitatively, we simulated cavities with $\delta r_{max}$ set to the 20\% reduction in form factor point of the $\ell = 38.75$ mm cavity across several different $\ell$ values. The results of this investigation are summarized in Figure \ref{fig:radii_variance_20_percent_reduction}. 
Again, the data points represent the mean value for the appropriate quantity across 10 simulated cavities, with the error bars reporting the standard deviation. We see that, indeed, the form factor increases as the coupling increases ($\ell$ decreases), a clear indication that the tolerance requirement is relaxing. The unloaded quality factor exhibits a slight decrease; however, this behavior is related to the decrease in the cavity's resonant frequency (see upper horizontal axis) as the cells are brought closer together.

In total, our simulations demonstrate that strong coupling between cells offers a significant benefit with respect to the manufacturing tolerances required to operate the system successfully. The required tolerance can be relaxed ($\delta r_{max}$ increased) further by increasing the coupling between adjacent cells.

To investigate the required assembly tolerance, we modified our cavity generation technique to produce beehives with imperfect rod positions. In particular, using a similar (but now multivariate) truncated normal distribution, we pulled $n$ random rod positions ($x$ and $y$-coordinates), each corresponding to one of the cells within the beehive. In our results below, we report position variation in terms of a maximum root-mean squared (RMS) position error $\delta d_{max}$. This quantity indicates that the standard deviations of the multivariate distribution were set to $(\sigma_x, \sigma_y) = \Big{(} \frac{\delta d_{max}}{\sqrt{2}}/2, \frac{\delta d_{max}}{\sqrt{2}}/2 \Big{)}$, while the truncation limits were set to $[x_{min}, x_{max}] = \Big{[} -\frac{\delta d_{max}}{\sqrt{2}},\frac{\delta d_{max}}{\sqrt{2}} \Big{]}$ and $[y_{min}, y_{max}] = \Big{[} -\frac{\delta d_{max}}{\sqrt{2}},\frac{\delta d_{max}}{\sqrt{2}} \Big{]}$. Again, we are assuming that the position errors are random independent variables up to a defined tolerance value. 

\begin{figure}
    \includegraphics[left]{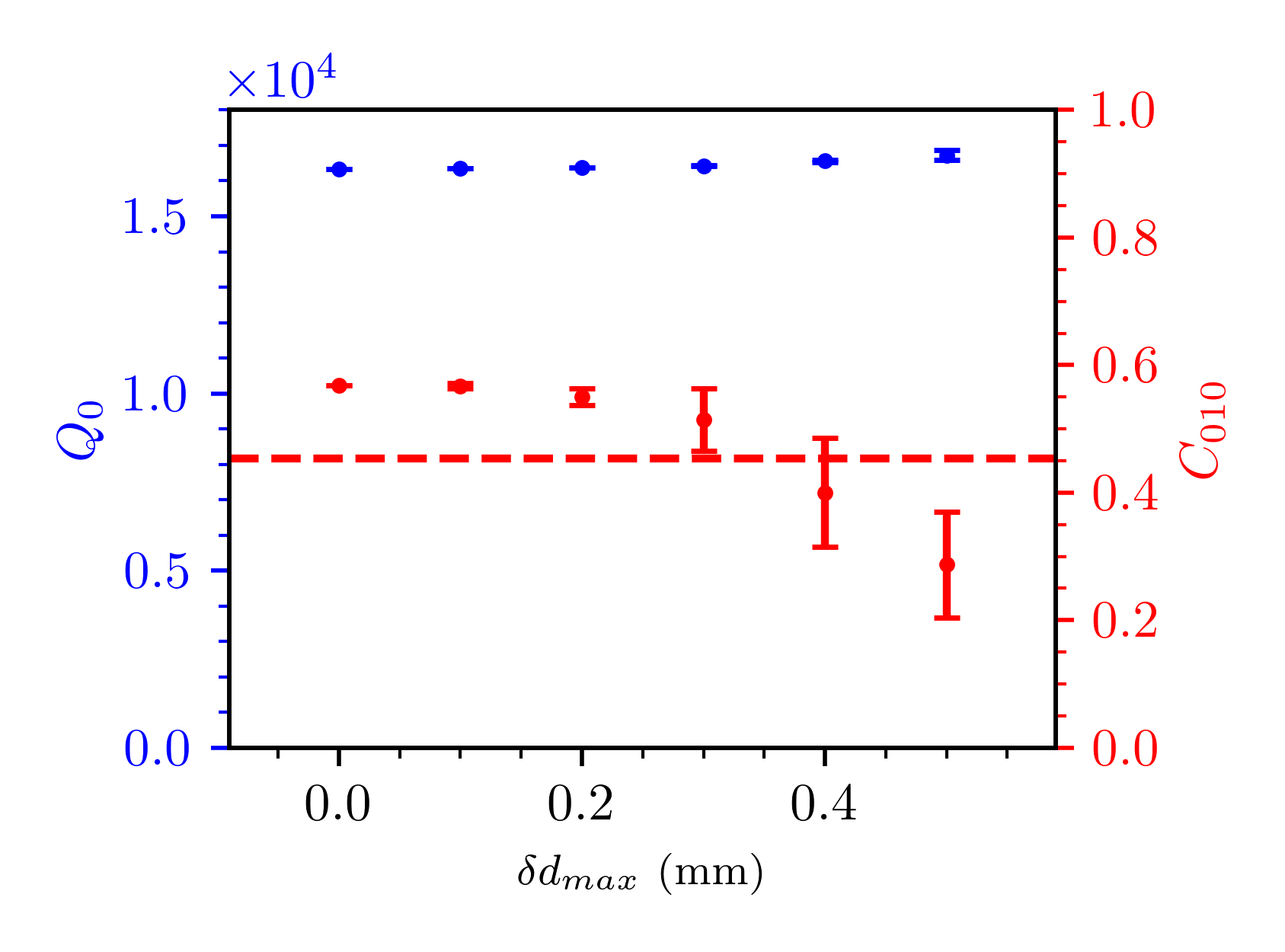}
    \caption{Unloaded quality factor $Q_0$ and form factor $C_{010}$ vs. RMS displacement variation $\delta d_{max}$ introduced to tuning rod positions. The cavity dimensions were $n = 169$, $\ell = 38.75$ mm, $R = 19.5$ mm, and $r= 7$ mm. Each data point is the average result of 10 different randomly generated cavities, with the error bars indicating the standard deviation. The dashed horizontal line indicates the point at which the form factor has been degraded by 20\%. }
    \label{fig:rod_displacement}
\end{figure}

In Figure \ref{fig:rod_displacement}, we show the effect of varying $\delta d_{max}$ on $Q_0$ and $C_{010}$ for a cavity with $R = 19.5$ mm, $r =7$ mm, and $\ell = 38.75$ mm.
As in our previous summary plots, each data point represents the average of 10 different randomly generated beehive cavities, and the error bars indicate the standard deviation of the 10 samples. Notice that $Q_0$ remains relatively stable; however, $C_{010}$ has fallen by 20\% (dashed red line) at $\delta d_{max} = 0.352$ mm. Form factor reduction is again due to detuning, which eventually results in decoherence between the cells. 

\begin{figure}
    \includegraphics[left]{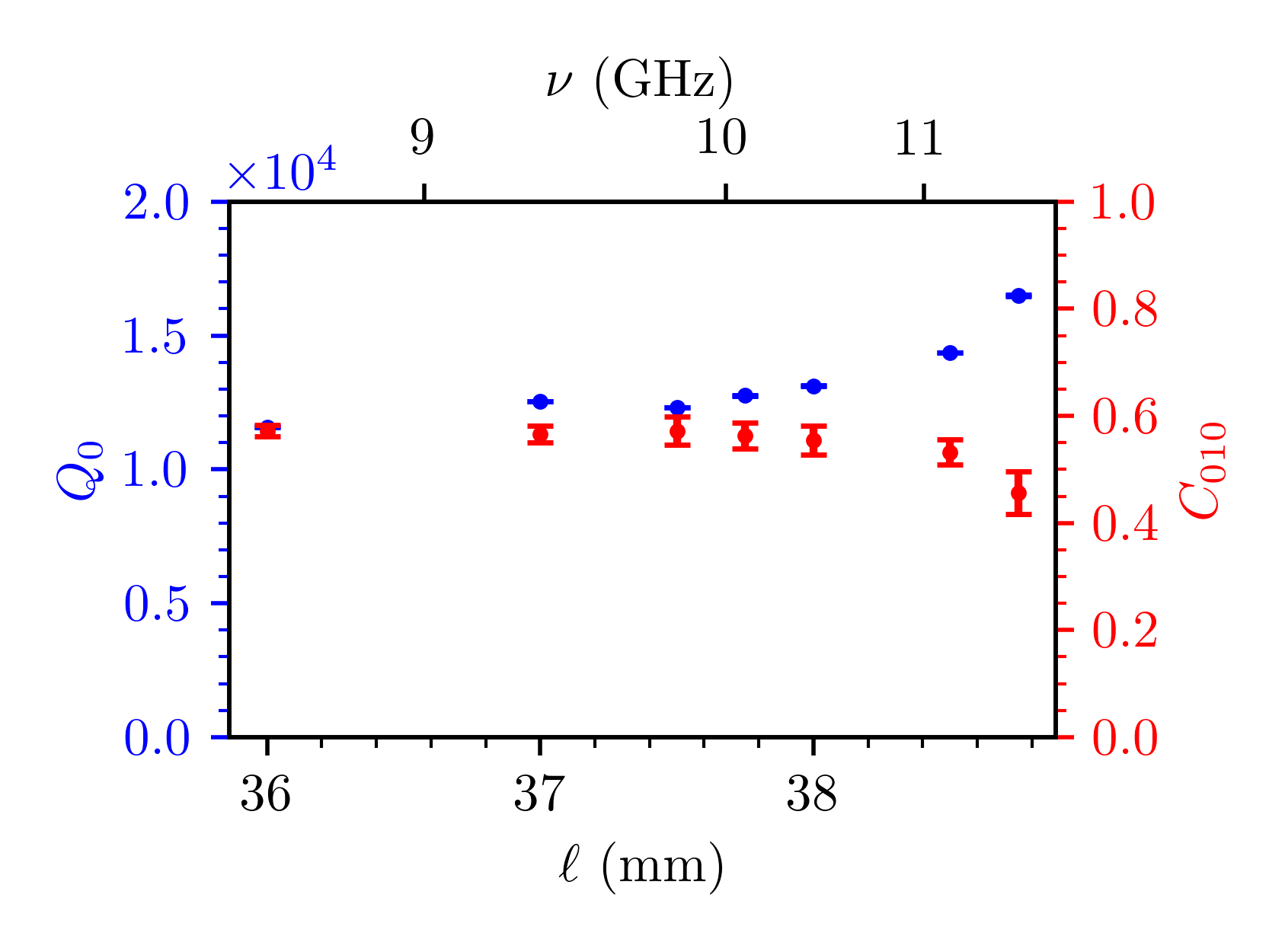}
    \caption{Unloaded quality factor $Q_0$ and form factor $C_{010}$ vs. cell separation $\ell$ for cavities with 0.352 mm of RMS displacement error. This displacement error results in a 20\% form factor reduction for a cavity with $\ell = 38.75$ mm. Other cavity dimensions were $R = 19.5$ mm, $r = 7$ mm, and $n =169$.}
    \label{fig:position_variance_change_coupling}
\end{figure}

Finally, like in the radii variance case, we also investigated the relationship between cell coupling (\textit{i.e.}, cell separation $\ell$) and the resilience of the system with respect to alignment errors. In Figure \ref{fig:position_variance_change_coupling}, we plot $Q_0$ and $C_{010}$ vs. cell separation when $\delta d_{max} = 0.352$ mm of RMS displacement error is injected onto the cell positions. 
As mentioned above, this particular amount of displacement corresponds to a 20\% decrease in form factor compared to an ideal cavity with identical dimensions. As is our custom, each data point is the average of 10 different randomly generated cavities, and the error bars represent the standard deviation of this set. Notice that for the same amount of RMS displacement error, the form factor increases for stronger couplings (smaller values of $\ell$) while the quality factor decreases. This, too, is qualitatively consistent with our theoretical exploration in part IB (\textit{i.e.}, increasing coupling increases the mode splitting and relaxes the tolerance requirements). 

Precision metrology techniques can routinely machine parts to within 0.05 mm (see, for example, \footnote{\url{https://www.mcmaster.com/products/rods/stainless-steel~/shape~rod-and-disc/diameter~7-mm/}}) and measure position accuracy to the order of 0.02 mm (see, for example, the Hexagon Absolute-Arm 85 series in \footnote{\url{https://willrich.com/wp-content/uploads/2014/09/Hexagon-MI-Absolute-Arm-2018-Brochure-.pdf}}). Our reported tolerances are within this range, demonstrating that the beehive cavity is resilient enough to be considered a good candidate for construction and tuning.

\subsection{3D Simulations}
We now turn our attention to the simulated behavior of realistically implemented beehive haloscopes in three dimensions. The 2D simulations explored above are really transverse cross-sections of an ideal 3D cavity. In particular, they represent the magnitude of the longitudinal ($z$-direction) electric field within a beehive haloscope assuming (a) the beehive is closed, with conducting surfaces (``endcaps") at the top and the bottom and (b) the tuning rods span the entire longitudinal direction (\textit{i.e.}, there are no gaps between the rods and the endcaps).  The 2D calculations are additionally limited because any modes with a varying $E_z$ are omitted.  Finally, the $E_{/\!/}=0$ boundary conditions at the endcaps imply all the eigenmodes that contain non-vanishing $E_x$ and $E_y$ components are artificially excluded. Hybridization effects with these spurious modes must be studied in 3D FEA in conjunction with experimental measurements.  

For these 3D simulations, we have chosen a base geometry of $n=7$, $\ell = 38$ mm, $R = 19.5$ mm, $r = 7$ mm, and $h=300$ mm. The tuning range  is similar to that of the $n=169$-cell cavity in our 2D sims (compare Figures \ref{fig:ideal_tuning_freq} and \ref{fig:3D_freq}). A beehive haloscope with $n=7$ cells is the simplest geometry with a complete ring of cells (ignoring the trivial case of an $n=1$-cell cavity, which technically forms the zeroth ring of any hive). Using the smallest possible number of cells is not only advantageous given limited computational resources but it also represents a feasible geometry for an initial prototype of the system. The results we present here should hold (at least qualitatively) in hives with larger $n$, where the volume of the system is greatly increased to enhance the scan rate of the instrument.

In Figures \ref{fig:3D_freq}-\ref{fig:3D_C010}, we show the frequency $\nu$, unloaded quality factor $Q_0$, and form factor $C_{010}$ of several different beehive geometries, all with $n=7$. For now, notice that the percent difference between the simulated eigenfrequency of an ideal 3D cavity and the simulated eigenfrequency of a 2D cavity constructed with the same geometry is 0.15\% on average across the tuning range. Likewise, the ideal 3D and 2D cases exhibit a percent difference of 1.8\% for unloaded quality factor and 1.4\% for form factor (again, we report percent difference values averaged across the tuning range). Knowing that there is strong quantitative agreement between results derived from a 2D geometry and results derived in 3D helps confirm the predictive nature of the 2D simulations we explored in section IIIA. 

Despite its correspondence with 2D simulations, the ideal beehive presents major practical challenges, especially related to tuning. Our scheme relies upon our ability to move all of the rods simultaneously in the $xy-$plane. Thus, we will likely need to connect them to a common platform outside the cavity (see Figure \ref{fig:concept}). This necessitates leaving the bottom of the cavity open, as the rods will need to extend downward and then have freedom to move around within their respective cells. Like in traditional halsocopes, it will also be necessary to leave a small gap between the tops of the tuning rods and upper endcap of the cavity. Otherwise, the two pieces would slide against one another, potentially damaging the cavity and rod surfaces and generating heat. The inner surfaces of axion haloscopes must be machined to high precision, and thus it is critical to avoid surface damage. Heat production would degrade the cryogenic performance of the system and thus the unloaded quality factor.

To explore how a more realistic cavity would perform in terms of eigenfrequency, unloaded quality factor, and form factor, we simulated a beehive cavity featuring an open bottom and tuning rods that a) extend downward from the bottom of the cavity and b) end 1 mm below the upper surface of the cavity. These results are labeled as ``Realistic 3D Cavity" in Figures \ref{fig:3D_freq} - \ref{fig:3D_C010}. Notice that we observe a drop in form factor across the entire tuning range and a noticeable decrease in unloaded quality factor when the rod displacement is greater than 7 mm, which corresponds to frequencies below approximately 8.5 GHz. 

\begin{figure}
    \centering
    \includegraphics[width=0.5\textwidth]{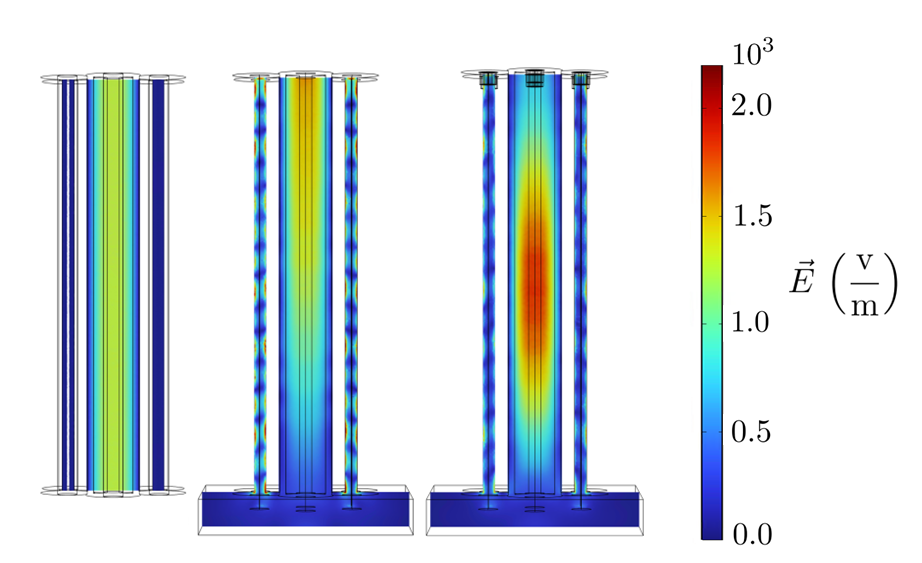}
    \caption{Cross sections of the simulated electric field strength of the TM$_{010}$ mode of an $n=7$ cavity constructed using ideal geometry (left), realistic geometry (center), and with $\lambda/4$ corrugation on top of the rods as a form factor loss mitigation technique (right). In each case the cavity is tuned ideally to $x=12$ mm. Note that these are all vacuum models, \textit{i.e.}, they simulate the electric field within the open space inside the cavity.}
    \label{fig:3D_sim_examples}
\end{figure}

\begin{figure}
    \includegraphics[left]{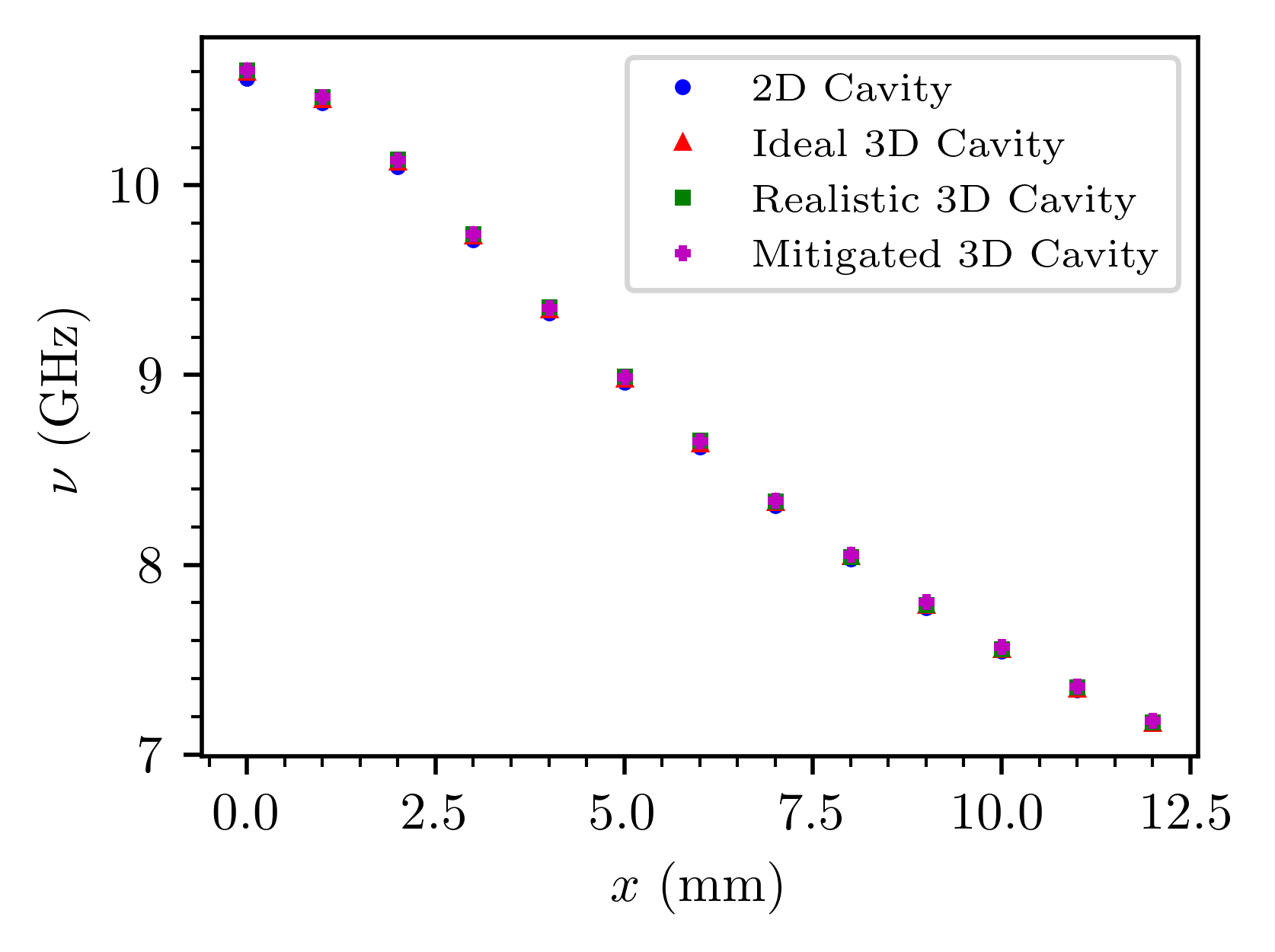}
    \caption{Resonant frequency $\nu$ versus tuning postion $x$ for a 2D cavity, an ideal 3D cavity, a realistic 3D cavity, and a mitigated 3D cavity, incorporating azimuthal corrugations on the top of the tuning rods. All cavities have $n=7$, $\ell = 38$ mm, $R=19.5$ mm, and $r=7$mm. All 3D cavities have $h=300$ mm, and the realistic and mitigated 3D cavities have a 1 mm gap between the tuning rod and the endcap. Notice that the cavities under consideration exhibit nearly identical resonant frequencies.}
    \label{fig:3D_freq}
\end{figure}

\begin{figure}
    \includegraphics[left]{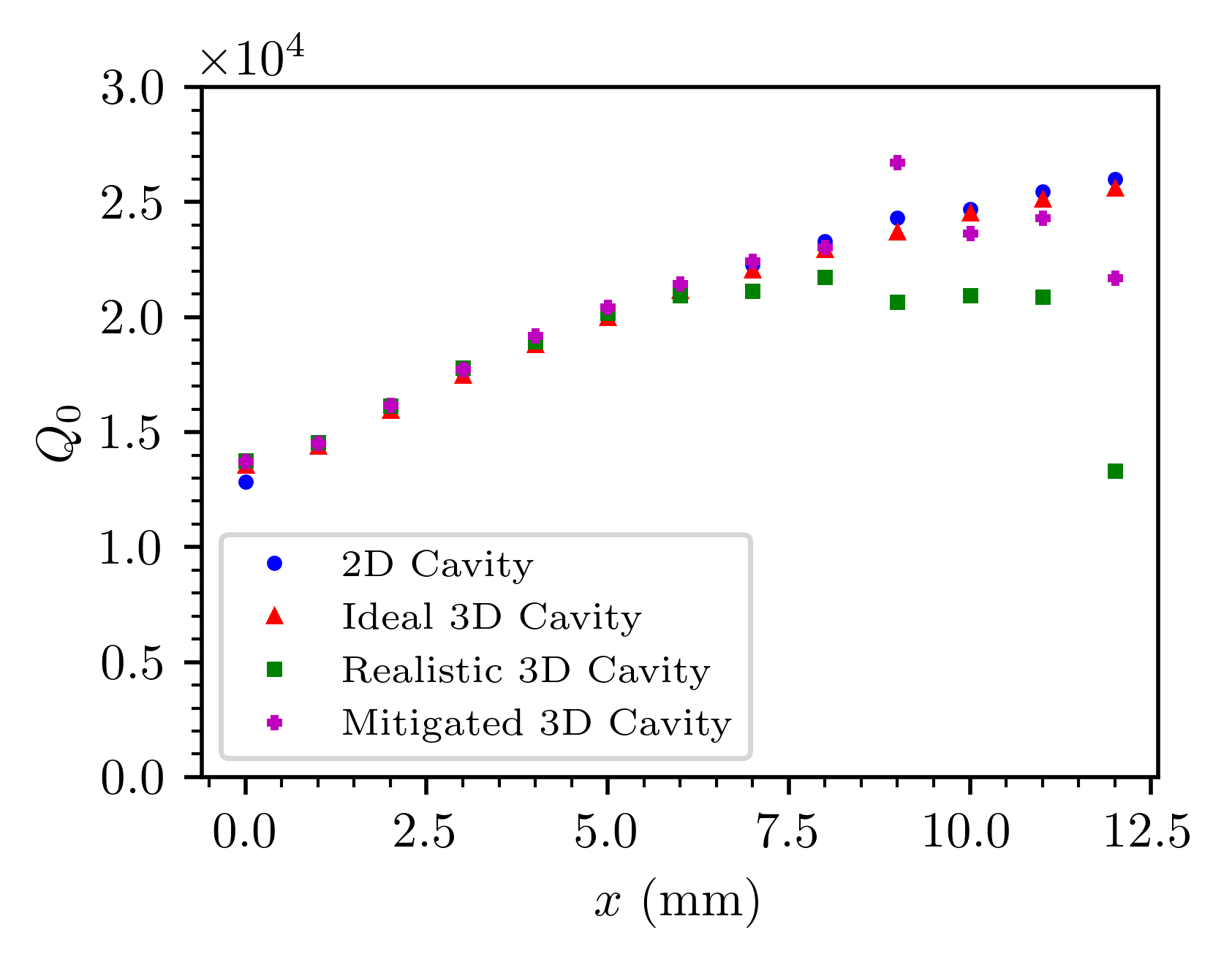}
    \caption{Unloaded quality factor $Q_0$ versus tuning postion $x$ for a 2D cavity, an ideal 3D cavity, a realistic 3D cavity, and a mitigated 3D cavity, incorporating azimuthal corrugations on the top of the tuning rods. All cavities have $n=7$, $\ell = 38$ mm, $R=19.5$ mm, and $r=7$mm. All 3D cavities have $h=300$ mm, and the realistic and mitigated 3D cavities have a 1 mm gap between the tuning rod and the endcap. Notice the partial recovery of quality factor achieved by introducing corrugations.}
    \label{fig:3D_Q0}
\end{figure}

\begin{figure}
    \includegraphics[left]{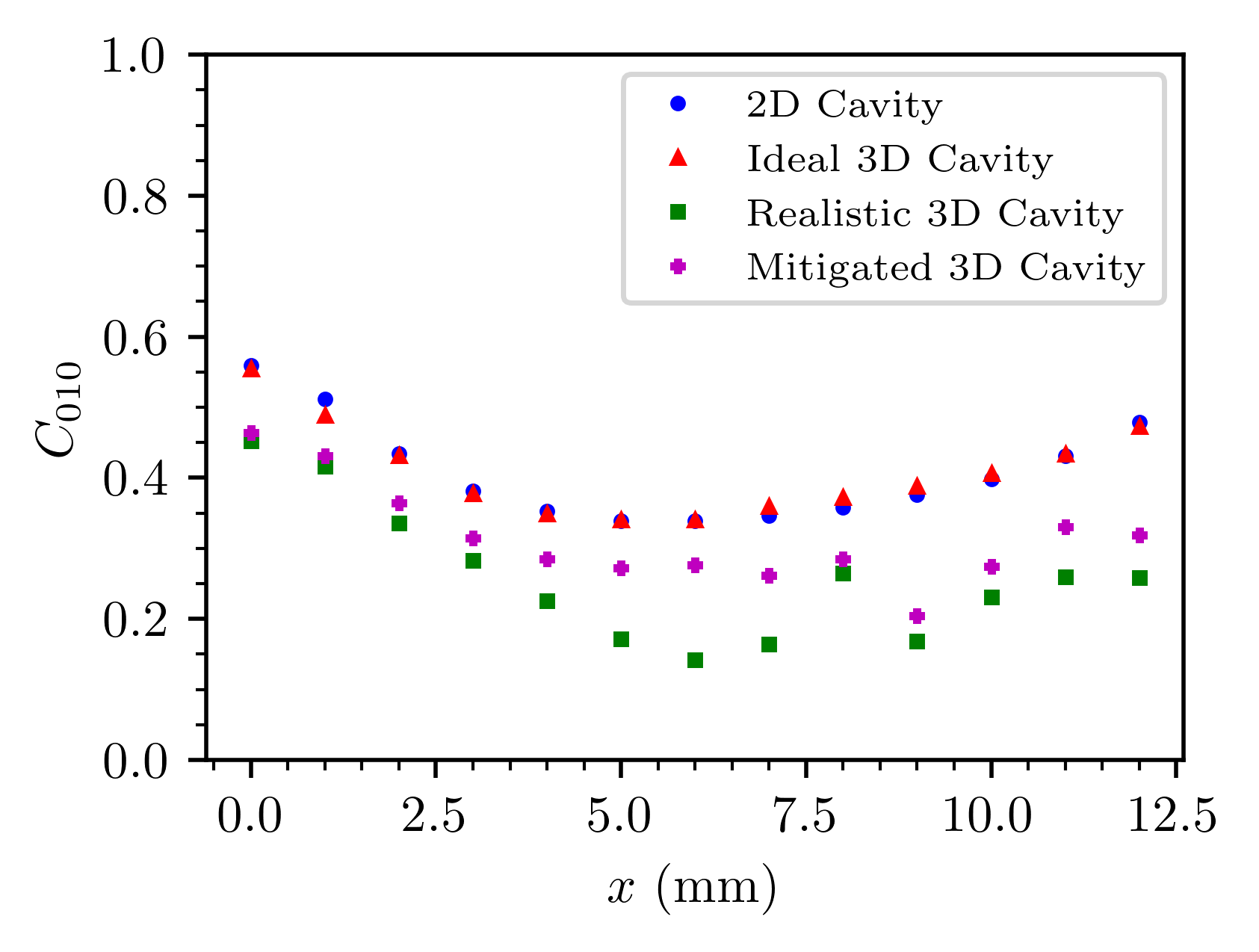}
    \caption{Form factor $C_{010}$ versus tuning postion $x$ for a 2D cavity, an ideal 3D cavity, a realistic 3D cavity, and a mitigated 3D cavity, incorporating azimuthal corrugations on the top of the tuning rods. All cavities have $n=7$, $\ell = 38$ mm, $R=19.5$ mm, and $r=7$mm. All 3D cavities have $h=300$ mm, and the realistic and mitigated 3D cavities have a 1 mm gap between the tuning rod and the endcap. Notice the partial recovery of form factor achieved by introducing corrugations.}
    \label{fig:3D_C010}
\end{figure}

The drop in form factor is attributable to a break in the longitudinal symmetry of the cavity's permittivity, which is introduced by the gap between the cavity endcap and the tuning rods \cite{stern2018}. In particular, capacitive effects become dominant within this region. In Figure \ref{fig:3D_sim_examples}, we show the consequences of this symmetry breaking. In the ideal cavity (left), longitudinal symmetry of the system ensures that pure TM$_{0n0}$ modes may form. However, when the rod-endcap gap is introduced to allow for tuning in the realistic cavity (middle), these modes experience perturbations, which may lead to mixing between TM, TE, and TEM modes across the tuning range and a corresponding degredation of the form factor. The relevant perturbations are visible in Figure \ref{fig:3D_sim_examples} as the ``ripples" that appear in the electric field magnitude.

One potential mitigation technique for this problem is the use of $\lambda/4$ corrugations to drive the mode away from the rod-endcap gap, increasing symmetry and reducing hybridization with TE and TEM modes. As noted in \cite{Kuo_2020} and demonstrated in \cite{dyson} for the single-wedge prototype cavity, the introduction of corrugations at a boundary inhibits the ability for the mode to exist within the immediate vicinity. We developed an additional set of simulations, labeled ``Mitigated 3D Cavity" in Figures \ref{fig:3D_freq} - \ref{fig:3D_C010}, in which we placed azimuthal corrugations 8.59 mm deep and 0.933 mm wide at the top of each tuning rod. The corrugation depth corresponds to $\lambda/4$ for the central frequency of the cavity's tuning range. With this corrugation in place, we observe a partial recovery of the form factor compared to the realistic 3D cavity case (see ``Mitigated 3D Cavity" in Figure \ref{fig:3D_C010}). From the perspective of the electric field, Figure \ref{fig:3D_sim_examples} demonstrates that the ripples which arose due to hybridization now have a reduced magnitude. We note that more optimization of the geometry and placement of these corrugations is likely needed. We plan to pursue this work as part of our efforts to develop a working prototype of the beehive haloscope.

Reduction in unloaded quality factor is mainly attributable to the large radiative loss that is introduced by leaving the bottom of the cavity open. Like in the  single-wedge prototype \cite{dyson}, it is likely possible to use additional $\lambda/4$ corrugations, positioned strategically near the cavity opening, to reduce these radiative losses. However, because the placement of the corrugations will depend heavily on the selected rod motion scheme, we defer an intensive study of this approach to a later date.

\begin{figure}
    \includegraphics[left]{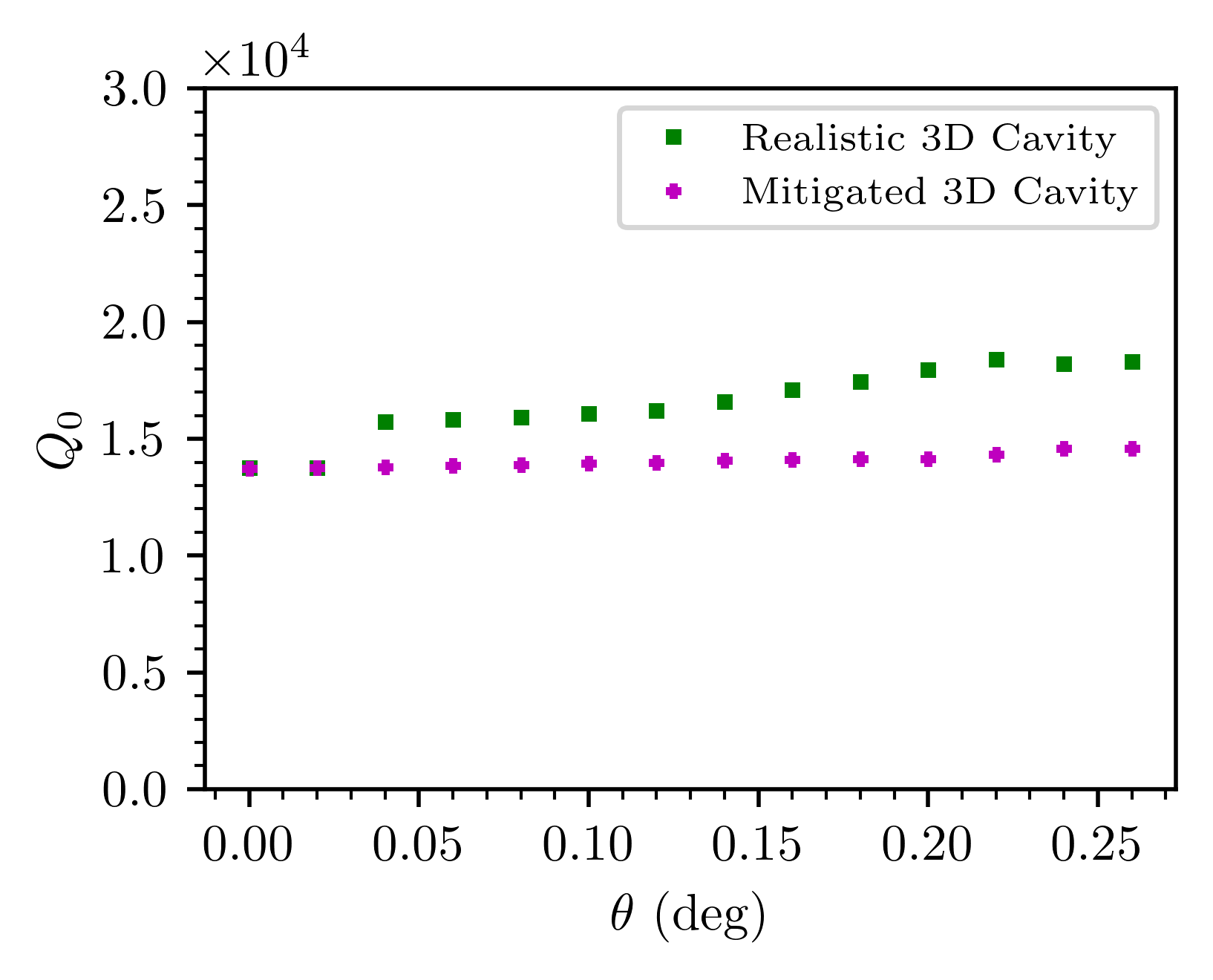}
    \caption{Unloaded quality factor $Q_0$ versus rod angle for the realistic and mitigated 3D cavities. Both cavities have $n=7$, $\ell = 38$ mm, $R=19.5$ mm, and $r=7$mm.}
    \label{fig:tilt_Q0}
\end{figure}

\begin{figure}
    \includegraphics[left]{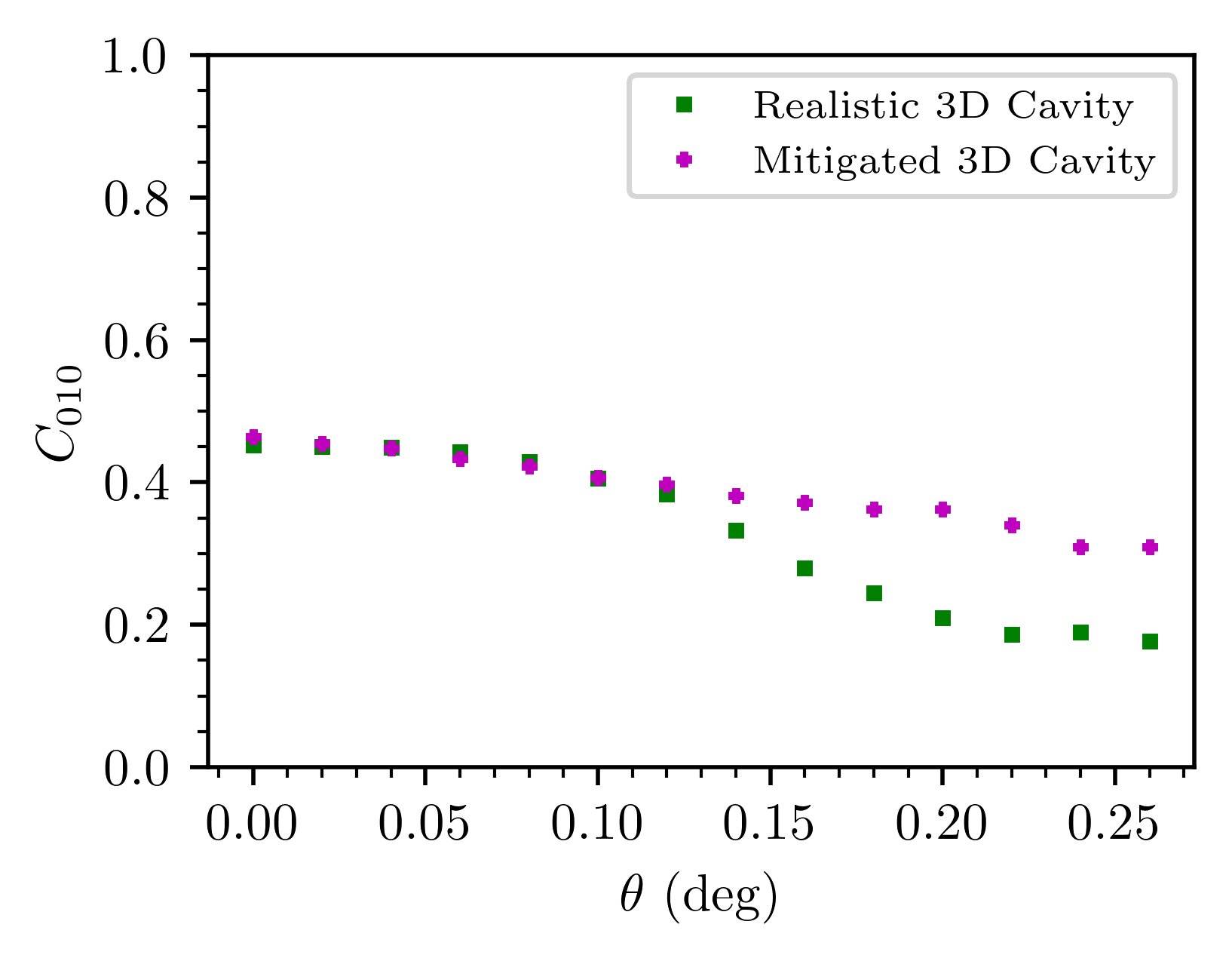}
    \caption{Form factor $C_{010}$ versus rod angle for the realistic and mitigated 3D cavities. Both cavities have $n=7$, $\ell = 38$ mm, $R=19.5$ mm, and $r=7$ mm. Notice that the mitigated cavity demonstrates a higher form factor at larger tilt angles.}
    \label{fig:tilt_C}
\end{figure}

In 2D, we studied form factor degradation that occurs due to error in the rod positions. In 3D, tilting of the rods can also degrade the form factor. In Figures \ref{fig:tilt_Q0} and \ref{fig:tilt_C}, we report the effect of rod tilt on the simulated unloaded quality factor and form factor.
We see an appreciable reduction in form factor as the tilt angle grows above $\sim0.1^\circ$ (see Realistic Cavity in Figure \ref{fig:tilt_C}). Furthermore, introducing azimulthal $\lambda/4$ corrugations to the top of the rods again helps to recover some of the form factor loss (see Mitigated Cavity in Figure \ref{fig:tilt_C}). 

\subsection{Implementation}

Since the building blocks of the beehive resonator are conventional coaxial cavities with essentially the same tuning mechanism, the new geometry retains many well known and appealing features of a cavity haloscope, including compatibility with solenoid magnets, straightforward fabrication, and well understood scaling of the quality factor. Nevertheless, there are a number of engineering issues that must be resolved in the future:  

\begin{itemize}
    \item Fabrication and metrology 
    
    The fabrication of the beehive cavity will involve drilling deep, overlapping bores into a block of metal and mounting an array of long rods on a platform with sufficient stiffness and precision. Capacitive or image-based metrology will have to be developed for the verification of the geometry. 
    
    \item Alignment and tuning 

    Similar to the approach taken by \cite{dyson}, the rod array will be placed relative to the hive using a positioning system with six degrees of freedom. Unlike the wedge design, the frequency tuning is achieved by small lateral movements. The positioning accuracy of the entire rod assembly must be better than $\lambda /2Q$ to allow for Nyquist sampling within the intended bandwidth \footnote{Note that here we consider position accuracy from the perspective of the tuning of the cavity through frequency space. Our previous simulations showing displacements $\mathcal{O}(100~\mu\textrm{m})$ served to set a limit on form factor breakdown. Achieving a position accuracy necessary for tuning will guarantee that the form factor remains high.}.
    
    \item Signal coupling

    One common challenge for many large-volume ($V\gg \lambda^3$) resonant haloscopes concerns the signal readout. Under-coupling to the resonator ensures a high $Q$ but leads to a weak signal output; over-coupling reduces the resonator $Q$ and therefore the axion-to-photon conversion power.  The maximum survey speed occurs when the coupling parameter $\beta$ is equal to two \cite{ALKENANY201711}, implying that the amount of signal power coming out of the coupling port is twice that dissipating on the cavity walls. Since the latter increases with the cavity volume, the coupling must also increase for a cavity with a large volume (and the same $Q$). This cannot be done by increasing the physical dimensions of a single port because doing so tends to split the eigenmode into a very lossy mode around the port and a high-$Q$ mode away from the port.  Most likely, the required slight over-coupling must be accomplished by summing the signal from a distributed array of probes throughout the cavity volume, as pointed out in \cite{Kuo_2020}. Despite this complication, we still expect the total number of ports to be much fewer than the number of cells. 
\end{itemize}

    A prototype beehive resonator is currently being developed to address these engineering issues.

\section{Discussion}\label{sec:discussion}
\subsection{Comparison to Other Geometries}
\begin{figure*}
    \centering
    \includegraphics[width=\textwidth]{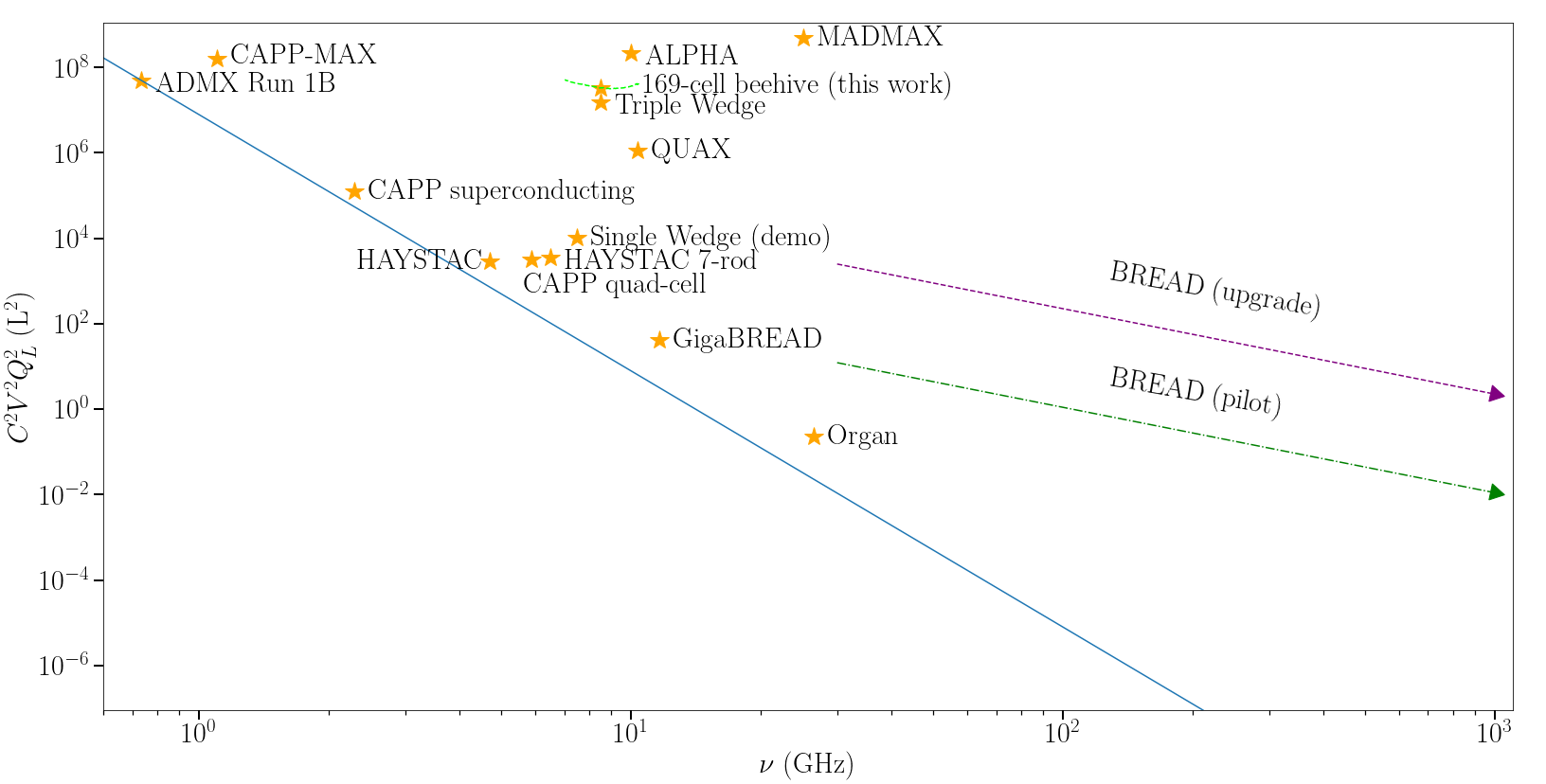}
    \caption{Comparison of various extant and proposed axion search experiments according to the $\frac{d\nu}{dt} \propto C^2 V^2 Q_L$ figure of merit. Relevant assumptions for each experiment are described in the text. Our proposal deviates from the expected $\frac{d\nu}{dt} \propto \nu^{-6}$ scaling for cavity haloscopes to achieve a highly competitive figure of merit.}
    \label{fig:fom_vs_freq}
\end{figure*}

\begin{figure*}
    \centering\includegraphics[width=\textwidth]{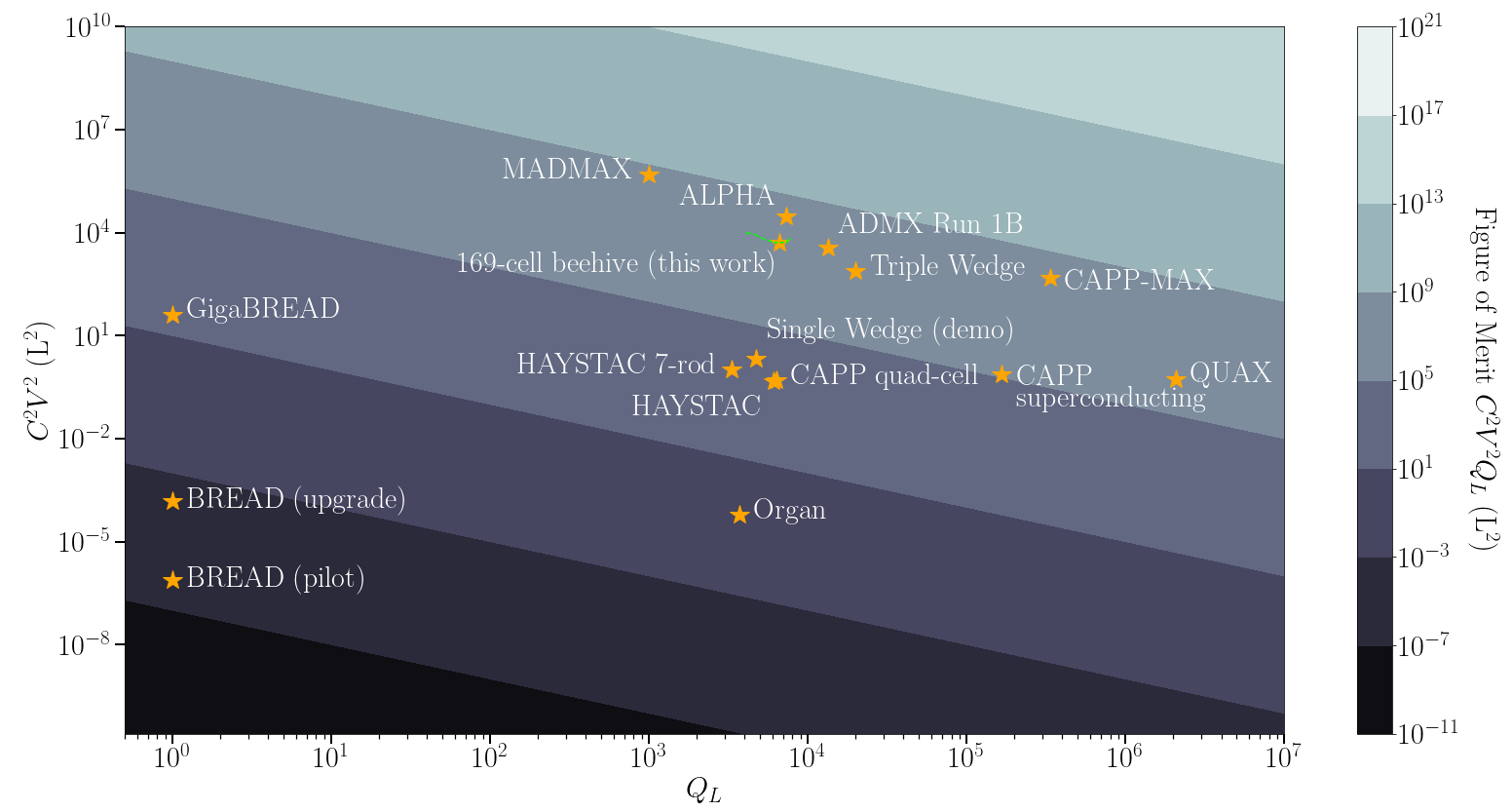}
    \caption{Comparison of various extant and proposed axion search experiments, highlighting individual contributions of effective volume and quality factor. Relevant assumptions for each experiment are described in the text. Note that our proposal's high figure of merit arises primarily due to improvement in effective volume rather than quality factor.}
    \label{fig:FOM}
\end{figure*}

The beehive design was conceived during the study of the 7 GHz thin-shell haloscope \cite{dyson}, which can be seen as two cells joined at the ends around the tuning wedge to form a single cavity.  We noticed that the fundamental TM$_{010}$ mode in the resulting cavity corresponds to the symmetric hybridization of the cell resonances and that the form factor is much more resilient to relative detuning of the two cells. As explained in Section IB, the resilience increases with the coupling strength between cells. This observation motivates the beehive design that seeks to maximize the interactions between the cells.  Comparing these two concepts, the thin-shell wedge design that consists of flat surfaces has the advantage in its ease of fabrication and metrology, while the collective oscillator behaviors in a beehive array significantly relax the mechanical tolerance requirements. 

We also wish to briefly comment on the apparent resemblance between the beehive cavity and the plasma haloscope that was proposed in \cite{plasma} and developed further in \cite{alpha}. In their current implementations, both schemes involve an array of elongated metallic elements running in parallel, a tuning scheme that shifts half of these elements laterally, and a resonant mode with $E$-fields parallel to the applied $B$-field. To the best of our knowledge, however, the two concepts rely on resonances in different regimes.  The beehive cells are resonant because of local standing waves between the rods and the cylindrical walls, while in the plasma haloscope the electrons in the metallic elements react resonantly to the waves scattered over the entire effective medium. We therefore conclude that the resemblance is only superficial without providing a detailed theoretical comparison. The beehive cavity also exhibits a passing resemblance to the HAYSTAC Collaboration's work with photonic band gap (PBG) structures \cite{PhotonicBandGap}. A cavity constructed out of PBGs can be designed to support only TM modes and thus eliminate mode crossings. Again, the similarities between the two schemes are superficial, as the PBG structures exists outside the main eigenmode and such cavity was not proposed to directly evade the $\nu^{-6}$ scaling in $d\nu/dt$. 

Next we compare the predicted performance of the proposed beehive haloscope to other extant and proposed axion search experiments. One natural method of comparison is using figure of merit $d\nu/dt \propto C^2V^2Q_L$. We note that the volume of the resonator in our scheme ultimately is limited only by the size of the magnet.  Thus, to develop a fair comparison, we present a cavity which roughly fills the cavity-reserved portion of the magnetized cold volume of the ADMX-EFR experiment ($\sim0.6$ m diameter, 1 m height). If we give our benchmark 169-cell cavity (with tuning properties as described in Figures \ref{fig:ideal_tuning_freq} and \ref{fig:ideal_tuning_Q0_and_C}) a height of 1 m, the maximum linear dimensions become $0.569~\textrm{m}\times0.499~\textrm{m{}}\times1~\textrm{m}$, falling within the ADMX limits. The overall active volume (as computed in COMSOL-RF) is 173 L. Note that in a real-world implementation, the width and depth of the beehive would likely need to be reduced slightly to allow space for cryogenic components (\textit{e.g.}, radiation shields).

In Figures \ref{fig:fom_vs_freq} and \ref{fig:FOM}, we apply the figure of merit to several operating and proposed axion search experiments. Traditional and experimental cavity halscopes (\textit{e.g.}, our work, thin-shell wedge cavities \cite{dyson}, ADMX \cite{ADMX1B_Analysis}, HAYSTAC \cite{HAYSTAC:2023cam}, CAPP-MAX \cite{ahn2024extensive} CAPP quad-cell \cite{jeong18}, CAPP superconducting \cite{CAPPSuperconductingPathfinder}, ORGAN \cite{ORGAN}, and QUAX \cite{QUAX}) have well-defined values for $C$, $Q_L$, and $V$. For experiments reporting only an unloaded quality factor (this work, HAYSTAC 7-rod \cite{Simanovskaia_2021}, CAPP experiments, and the thin shells), we assumed a coupling of $\beta = 2$ since this will maximize their scan rate \cite{ALKENANY201711}. Since BREAD \cite{bread} and GigaBREAD \cite{Knirck:2023jpu} are dish antennas, we approximate their equivalent volume using $V = A\lambda/2$, where $A$ is the area of the dish and $\lambda$ is the operating wavelength. We set their $Q_L$ and $C$ values to 1 to aid in comparison to resonant cavities. We plot the two BREAD experiments (pilot and upgrade) as curves since the scheme targets a broad frequency range beginning at $\nu = 30$ GHz. MADMAX \cite{MADMAX} is a dielectric haloscope, which uses discontinuities in the non-propagating axion-induced electric field at vacuum-dielectric boundaries to produce propagating electric fields. These propagating fields can then be brought into constructive interference and detected with a receiver chain. The MADMAX collaboration reports formulae for examining their experiment from the perspective of a resonant haloscope \cite{madmax_17}. While their final experimental geometry is still in flux, conversations with their team reveal that $C\approx0.7$, $Q_L\approx1000$, and $V\approx1000$ L are sufficiently representative of their plans. For plasma haloscope ALPHA, we used the $Q_0$ predicted for a cryogenic experiment operating at 10 GHz (see Figure 4 in \cite{alpha}). We once again assumed $\beta = 2$ when converting this value to $Q_L$.

Figure \ref{fig:fom_vs_freq} plots the entire figure of merit against frequency, highlighting the challenge of maintaining a robust scan rate at $\mathcal{O}(>10~\textrm{GHz})$ frequencies. The solid blue line represents $d\nu/dt \propto \nu^{-6}$ scaling of ADMX Run 1B; we introduce this as a visual representation of the expected degradation in scan rate that occurs when building a traditional haloscope cavity for higher frequencies. We see a collection of experiments lying slightly above this line. These searches generally obey the expected scaling, with slightly improved performance due to notable technological improvements (\textit{e.g.,} CAPP's use of superconducting tape to coat the cavity walls and improve $Q$). Our 169-cell beehive falls within a second cluster, which includes dielectric and plasma haloscopes, as well as a full-scale proposal for thin shell haloscopes (see Triple Wedge). These experiments collectively extend the scan rate of cavity halsocopes operating in the $\mathcal{O}(100$s MHz$-1$ GHz) regime into the $\mathcal{O}(1-10$s GHz) regime.

In Figure \ref{fig:FOM} we place the loaded quality factor $Q_L$ on the horizontal axis, the square of the effective volume $C^2 V^2$ on the vertical axis, and plot contour bands corresponding to orders of magnitude in the figure of merit. Under this representation, faster scanning experiments are located in the lighter contour bands. We can also clearly see the individual contributions of effective volume and quality factor to the scan rate. Our proposal is again distinguished from traditional cavities at high frequency. It is clear that the primary advantage is due to improvement of the system's effective volume, as the quality factor $Q_L$ is of the same order of magnitude.

\begin{figure*}
    \includegraphics[width=0.75\textwidth]{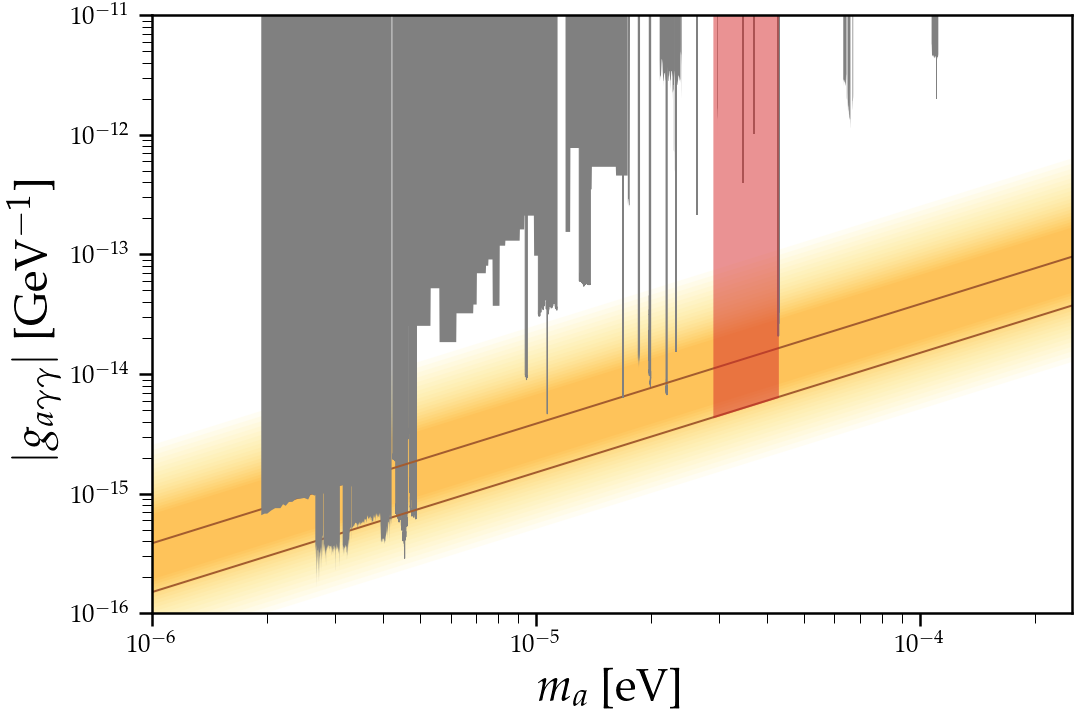}
    \caption{Axion parameter space (shown in red) accessible to a beehive haloscope with geometric parameters $n=169$, $\ell = 38$ mm, $R=19.5$ mm, $r=7$ mm, and $h = 1$ m. As explained in the text, expected scan times to cover this range are 0.25 years to KSVZ level and 0.93 years to DFSZ level. Gray regions indicated parameter space that has already been excluded by other experiments. The tan region indicates a band of uncertainty in $g_{a \gamma \gamma}$ for the KSVZ and DFSZ models. This plot was generated using software provided by \cite{LimitPlotCode}.}
    \label{fig:parameter_space_coverage}
\end{figure*}

In principle, the beehive haloscope can be made to fill the entire magnetized volume of a solenoid. The $n=169$-cell case explored in-depth throughout this paper approximates a reasonable magnetized volume-filling cavity if a height of $\sim 1$ m is assumed. In Figure \ref{fig:parameter_space_coverage}, we show the potential parameter space (down to DFSZ sensitivity) available to this cavity. To estimate the time required to scan this space, we note that the total digitization time is given by 
\begin{align*}
    T_{dig} = \int_{\nu_i}^{\nu_f} d\nu \Big{(} \frac{d\nu}{dt} \Big{)}^{-1}, 
\end{align*}
where $\frac{d\nu}{dt}$ is the instantaneous scan rate and $\nu_i$ and $\nu_f$ are the initial and final frequencies accessible to the cavity, respectively. If we also assume an additional time $T_{anc} = \int_{v_i}^{v_f} d\nu C$ for ancillary measurements and procedures (\textit{e.g.}, tuning motion, antenna coupling, cavity reflection and transmission measurements, \textit{etc.}) and an experiment efficiency of $\eta$, the total scan time $T_{scan}$ becomes
\begin{align*}
    T_{scan} = \frac{T_{dig} + T_{anc}}{\eta}.
\end{align*}
This equation, of course, ignores rescan time, which is heavily dependent upon the number of axion candidates, and, consequently, the specifics of the analysis code which flags them. If we assume the values summarized in Table \ref{tab:scan_time_params}, we can cover the $n=169$-cell hive's entire frequency range to KSVZ level in 0.25 years. Covering the same frequency range to DFSZ level with the same experiment parameters would require 0.93 years of scan time. Additional parameter space within the post-inflationary regime can be explored by manufacturing beehive haloscopes with different dimensions.
\begin{table*}[]
\centering
\caption{Parameters used to estimate scan time $T_{scan}$.}
\begin{tabular}{|c|c|l|}
\hline
\textbf{Parameter} & \textbf{Estimated Value} & \multicolumn{1}{c|}{\textbf{Justification}}                                                        \\ \hline
$[\nu_i, \nu_f]$    & [7.03 GHz, 10.4 GHz]      & Set by geometry of simulated beehive haloscope \\ \hline
$V$                 & 173 L                     & Approximately magnetized volume-filling implementation of the beehive haloscope scheme \\ \hline
$B_z$               & 10 T                     & Typical order of magnitude for the magnetic field in axion search experiments \\ \hline
$T_{sys}$         & 120 mK                   & Possible system temperature with a quantum limited flux-driven JPA \cite{Kutlu_2021}           \\ \hline
$\beta$            & 2                        & Antenna coupling which maximizes scan rate \cite{ALKENANY201711}                             \\ \hline
SNR                & 3.5                      & Typical SNR required by axion data analysis procedures \cite{ADMX1B_Analysis}                 \\ \hline
$\rho_{DM}$        & 0.45                     & Typical assumption for the local dark matter density \cite{ADMX1B_Analysis}                   \\ \hline
$T_{anc}$ / meas.          & 20 s                    & Typical time for ancillary measurements per digitization for ADMX \cite{ADMX1B_Analysis}             \\ \hline
$\eta$             & 0.8                      & Reasonable experiment efficiency                                 \\ \hline
\end{tabular}
\label{tab:scan_time_params}
\end{table*}

\section{Conclusion}
In this paper, we proposed and simulated a novel haloscope geometry inspired by beehives and targeting parameter space consistent with post-inflationary axions (mass of 20-200 $\mu$eV down to DFSZ senstivity). Combining elements of thin-shell and multi-cavity haloscopes and characterized by strong coupling between a large number of axially-aligned cylindrical cells, these ``beehive" haloscopes have a tuning range that is approximately $\pm 10\%$ of their central frequency and are inherently scalable. In other words, the volume of the resonator can be increased arbitrarily (up to the maximum volume permitted by the search experiment's magnetized volume) by increasing the number of cells. 2D simulations show that the quality and form factors see minimal degradation during such scaling. 

Strong coupling between adjacent cells significantly reduces the required machining and assembly tolerances of the system. In particular, our 2D simulations revealed that beehive cavities can be reliably tuned as long as machining tolerances are held to $\sim0.039$ mm and alignment tolerances are held to $\sim0.352$ mm. These represent challenging yet feasible constraints given current metrology tools and techniques.

3D simulations shed light on behavior of the system not captured in 2D. In particular, we observed that simulations of realistic cavities exhibit lower quality and form factors than simulations of ideal (and 2D) cavities due to radiative losses (for quality factor) and breaking of the cavity's longitudinal symmetry (for form factor). However, emerging mitigation techniques, including the introduction of strategically-placed $\lambda/4$ corrugations to manipulate the position of the electric field within the cavity, show potential to help mitigate these parameter reductions. 3D simulations also contributed to our understanding of assembly tolerance, in particular showing that rod tilt must be held to under 0.1$^\circ$ for cavities without $\lambda/4$ corrugation. Substantial form factor persists through a rod tilt of 0.25$^\circ$ when corrugations are introduced.

Assessment of the potential performance of the beehive halscope revealed that it has the potential to extend the scan rate of ADMX's flagship $\mathcal{O}$(1 GHz) experiment into the few GHz range. The experiment is also competitive with other proposed high-frequency axion searches, including other cavity haloscopes, dielectric haloscopes, plasma haloscopes, and dish antennas. Using reasonable assumptions of experiment parameters, we estimate that an example beehive haloscope with $V = 173$ L ($n = 169$) targeting the 7.03-10.4 GHz range can reach DFSZ sensitivity in $\sim1$ year.

\FloatBarrier

\section*{Acknowledgements}

The paper is based upon work supported by the National Science Foundation under Grant No. 2209576 and a KIPAC Innovation Award. MOW acknowledges support from the KIPAC Giddings Fellowship.

We are extremely grateful for numerous discussions with Chelsea Bartram, Taj Dyson, and Gray Rybka concerning the beehive haloscope concept and Stefan Knirck concerning accurate comparison to MADMAX.  We thank Junu Jeong for providing his COMSOL-RF implementation of the external current density induced by the axion-photon conversion. 

\bibliography{references}

\end{document}